%% file: main.tex
\documentclass{article} 
\usepackage{iclr2026_conference,times}
\usepackage{graphicx} 
\usepackage{amsmath}
\usepackage{float}
\usepackage{algorithm}
\usepackage{algpseudocode}
\usepackage{booktabs}
\usepackage{setspace}
\usepackage{amsthm}
\newtheorem{theorem}{Theorem}[section]
\newtheorem{lemma}[theorem]{Lemma}

\usepackage{xspace}
\usepackage{wrapfig} 
\usepackage{longfigure}
\input{math_commands}

\newcommand{\method}{InputDSA }
\newcommand{\methodshort}{iDSA}

\usepackage{hyperref}
\usepackage{url}

\iclrpreprintcopy

\title{\method: Demixing then comparing recurrent and externally driven dynamics
}

\author{
\textbf{\hspace{-2pt}Ann Huang}$^{1,2,3,\ast}$,
\textbf{Mitchell Ostrow}$^{4,\ast}$,
\textbf{Satpreet H. Singh}$^{2,3}$, \\
\textbf{Leo Kozachkov}$^{5}$,
\textbf{Ila Fiete}$^{4}$, 
\textbf{Kanaka Rajan}$^{2,3}$\\[4pt]
$^{1}$Harvard University \quad
$^{2}$Harvard Medical School \quad
$^{3}$Kempner Institute\\
$^{4}$Massachusetts Institute of Technology \quad
$^{5}$Brown University\\[4pt]
\texttt{\{annhuang@g.harvard, ostrow@mit\}.edu}\\[3pt]
$^{\ast}$Equal contribution
}

\begin{document}

\maketitle

\begin{abstract}
In control problems and basic scientific modeling, it is important to compare observations with dynamical simulations. 
For example, comparing two neural systems can shed light on the nature of emergent computations in the brain and deep neural networks. Recently, \citep{ostrow2024beyond} introduced Dynamical Similarity Analysis (DSA), a method to measure the similarity of two systems based on their recurrent dynamics rather than geometry or topology. However, DSA does not consider how inputs affect the dynamics, meaning that two similar systems, if driven differently, may be classified as different. Because real-world dynamical systems are rarely autonomous, it is important to account for the effects of input drive. To this end, we introduce a novel metric for comparing both intrinsic (recurrent) and input-driven dynamics, called \method (\methodshort). \method  extends the DSA framework by estimating and comparing both input and intrinsic dynamic operators using a variant of Dynamic Mode Decomposition with control (DMDc) based on subspace identification. We demonstrate that \method can successfully compare partially observed, input-driven systems from noisy data. We show that when the true inputs are unknown, surrogate inputs can be substituted without a major deterioration in similarity estimates. We apply \method on Recurrent Neural Networks (RNNs) trained with Deep Reinforcement Learning, identifying that high-performing networks are dynamically similar to one another, while low-performing networks are more diverse. Lastly, we apply \method to neural data recorded from rats performing a cognitive task, demonstrating that it identifies a transition from input-driven evidence accumulation to intrinsically-driven decision-making. Our work demonstrates that \method is a robust and efficient method for comparing intrinsic dynamics and the effect of external input on dynamical systems \footnote{Code is available at \href{https://github.com/mitchellostrow/DSA}{https://github.com/mitchellostrow/DSA}}.
\end{abstract}

\section{Introduction}

Identifying that two seemingly disparate complex systems have the same underlying structure is a widespread objective across many scientific fields, including deep learning \citep{huh2024platonicrepresentationhypothesis}, computational and systems neuroscience \citep{yamins2014performance,aldarondo2024virtual,prinz_similar_2004}, and physics \citep{hohenbergcritical,feigenbaum1978quantitative}.
One common approach to characterizing the similarity of two systems (e.g., brains, minds, computational models, or physical objects) is to compare the geometry of their states. Well-known methods to do so are Representational Similarity Analysis, Centered Kernel Alignment, Procrustes Analysis, Canonical Correlation Analysis, and Pearson Correlation  \citep{Kriegeskorte2008RSA, kornblith_similarity_2019, williams2022generalizedshapemetricsneural, gallego_cortical_2018,raghu_svcca_2017, schrimpf_brain-score_2018}. Neural networks can also be characterized by the topology of their activations, \citep{chaudhuri2019intrinsic, gardner2022toroidal,Trsa}, a more invariant measure than geometry, which depends on the particular sampling of neurons from the network. However, common to all is that they do not capture similarity in temporal dynamics \citep{galgali_residual_2023,maheswaranathan2019universality,ostrow2024beyond}. 

Metrics such as Dynamical Similarity Analysis (DSA, \citealt{ostrow2024beyond}) offer an important complementary lens to structure characterization, by proposing a similarity metric on the level of dynamics. DSA provides a linear, efficient and theoretically grounded dynamical similarity metric that has been successfully applied to recurrent network dynamics, training dynamics, and biological neural data \citep{redman2024not,huang2025measuring,codol2024brain,guilhot2024dynamical,versteeg2025computation,lazzari2025multitasking}. Briefly, DSA embeds nonlinear dynamics into a high-dimensional space and estimates a linear state-transition operator from observed trajectories, which is then compared across systems. Recent work introduced other methods for dynamics comparison \citep{redman2024,VermaniNassarJeonDowlingPark2024MetaDynamicalStateSpace,cotler2023analyzingpopulationsneuralnetworks,gosztolai_marble_2025,ChenVedovatiBraverChing2024DFORM,nejatbakhsh2024comparing} based on other computational techniques such as neural networks and shape metrics. Notably, none of these methods consider the effect of external input.

In neuroscientific settings such as central pattern generators or working memory circuits, dynamics may be treated as approximately autonomous \citep{MarderBucher2001CPG, Grillner2006NetworksInMotion, Kiehn2016SpinalCircuits, FusterAlexander1971ShortTermMemory, Funahashi1989MnemonicCoding, GoldmanRakic1995CellularBasisWM, Compte2000SpatialWM, Wang1999SynapticPersistentActivity}. Prior methods work well for comparisons in these settings. However, when activity is the result of both intrinsic dynamics and input drive, comparisons can be confounded by inputs. Most systems of interest in neuroscience and machine learning are non-autonomous, receiving sensory signals or communication from other subsystems \citep{eisen2025characterizing}. They are driven by complex external inputs and can receive observations that are contingent on the systems' outputs \citep{madhav2020synergy,kao2019neuroscience,RajanEtAl2010}.

Despite the ubiquity of input, current dynamical similarity methods ignore input-driven dynamics and do not incorporate estimation of how inputs affect states. To bridge this gap, we introduce \method (\methodshort), a method that disentangles intrinsic dynamics from input-drive, thereby enabling joint or separate metric comparisons of input-driven and intrinsic dynamics. \method extends the DSA framework by explicitly estimating both the intrinsic (state-transition) operator and the input-to-state mapping, which not only defines a new notion of similarity that incorporates the effect of inputs, but also in turn improves estimation of the intrinsic operator.



\begin{wrapfigure}{r}{0.45\textwidth}
    \centering
    \includegraphics[width=0.45\textwidth]{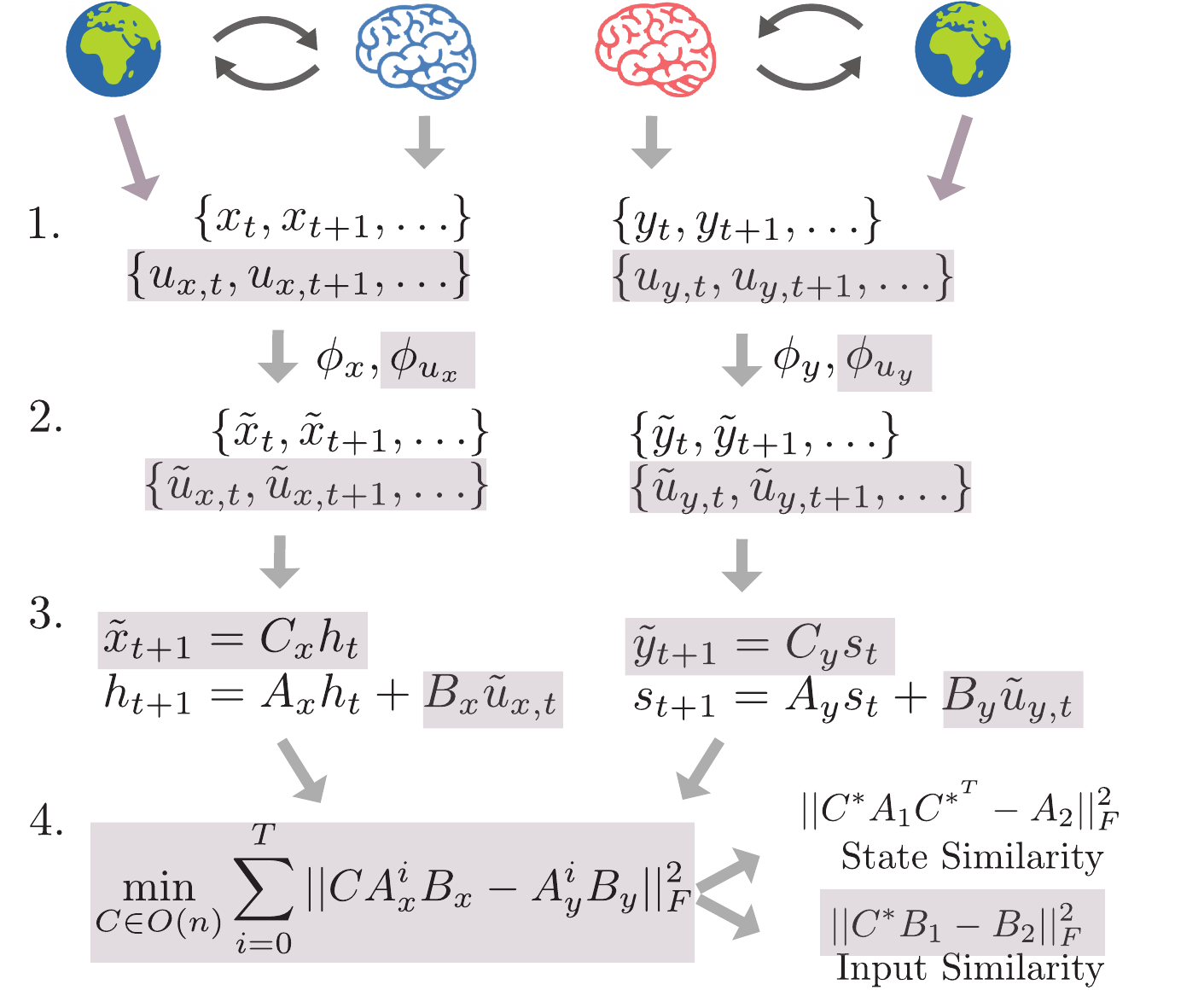}
    \caption{\textbf{\method schematic} (1), state and input data are collected from two systems. (2) data are embedded in a high-dimensional space (3) linear state-space models are fit to the data (4) Controllability, state, and input similarity are computed on learned state-space models. Gray indicates extensions from DSA.}
    \label{fig:fig1}
    \vspace{-18mm}
\end{wrapfigure}
 
\paragraph{Contributions} We extend DSA to non-autonomous systems that are driven by external input, which we call \method. To do so, we develop a novel similarity metric and variant of the dynamic mode decomposition (DMD), demonstrating that they can together provide complementary insights on both intrinsic as well as input-driven dynamical similarity. We demonstrate \method first on systems with known ground truth. We next show that similarity scores can be robust to surrogate or noisy inputs, provided that they have sufficient similarity to the real inputs. Finally, we apply \method to two datasets: RNNs trained with Reinforcement Learning, and neural population data (spiking) from rats performing a sensory decision-making task. We show that \method distinguishes high- from low-performing models and reveals how dynamics reorganize across different task periods.

\section{Methods}

\subsection{Dynamical Similarity Analysis (DSA)}

In dynamical systems, a key notion of similarity is called topological conjugacy: the existence of a homeomorphism that maps trajectories of one system onto those of another.  When two systems are conjugate, they have the same qualitative structure, including the same number and type of fixed points. Given two dynamical systems $f: X \rightarrow X$ and $g: Y \rightarrow Y$ with mapping $\phi : X \rightarrow Y$, (semi-) conjugacy is defined as:
\begin{equation}
g \circ \phi = \phi \circ f \label{eqn_conj}
\end{equation}
The existence of such a mapping entails a one-to-one alignment between topological features of each system such as invariant manifolds. Note that this is not geometric because distances and angles are not necessarily preserved under this mapping. In general, such a function can be arbitrarily complex, which can make searching for the true conjugacy map challenging in all but the simplest settings. DSA attempts to circumvent the optimization problem by approximating the Koopman Operator, which linearizes nonlinear dynamical systems via high-dimensional embeddings \citep{Koopman1931,budisic_applied_2012}. In the linear space, conjugacy maps are linear and therefore easier to identify. The methodology of DSA is therefore as follows: First, approximate your systems as linear in some high-dimensional space, yielding dynamics models $x_{t+1} = Ax_t$ . Then similarity is defined on the linear operators using the following metrics:
\begin{equation}\label{eq:DSA}
\text{DSA}(A_1,A_2)  := \, \min_{C \in O(n) } ||CA_1C^T - A_2||_F 
\end{equation}
\begin{equation}\label{eq: RedmanDSA}
\text{DSA}(\Lambda_1,\Lambda_2)  := \, \min_{P \in \Pi(n) } ||P\Lambda_1P^T - \Lambda_2||_F 
\end{equation}
Where $\Lambda_i$ is the eigenvalue matrix of $A_i$, and $O(n), \Pi(n)$ the groups of $n \times n$ dimensional orthogonal and permutation matrices. The latter metric was introduced by \citep{redman2024} and is a special case of the former \citep{ostrow2024beyond}. These metrics are reminiscent of Procrustes Analysis, which seeks an orthogonal transformation to align two data matrices, hence \cite{ostrow2024beyond} termed the first one Procrustes Analysis over Vector Fields. The latter metric is inspired from Koopman Operator Theory based on the relationships between Koopman Operators of conjugate systems \citep{budisic_applied_2012}. Other notions of similarity on the Koopman Operator are defined in \citep{mezic_comparison_2004,mezic_comparison_2016}.

\subsection{\method}

Inspired by DSA's approach for autonomous systems, consider two \textit{linear} dynamical systems
\begin{equation}\label{eq:linear_dynamics}
\dot{x} = A_1 x + B_1 u(t) \qquad \dot{y} = A_2 y + B_2 u(t).    
\end{equation}
A key feature of non-autonomous systems is their \textit{controllability}: the ability for an input sequence to drive the state to arbitrary points in finite time. In linear systems, this is encoded in the T-step controllability matrix (with T typically taken as the dimension of the system):
\begin{align}
    K_1(T) = \begin{pmatrix} B_1 & A_1B_1 & A_1^2B_1 & \dots & A_1^{T-1}B_1 \end{pmatrix}
\end{align}
and its corresponding Gramian, which encodes the geometry of controllability. 
\begin{align}
    W_c(T)= K_1K_1^T =  \sum_{i=0}^{T} A_1^iB_1(A_1^iB_1)^T
\end{align}
Intuitively, directions with small eigenvalues are easier to control, because they are more responsive to the effect of input. Controllability, as measured by the eigenvalues of the Gramian, is only preserved under orthogonal transformations between state spaces: 
\begin{align}
y = Cx \implies A_1 = CA_2C^T, B_1 = CB_2 \quad K_1 = CK_2
\end{align}
This motivates our proposed dissimilarity metric, which extends DSA:  
\begin{equation}\label{eq: \method}
\text{\method}(A_1,A_2,B_1,B_2,T) = \min_{\substack{C \in O(n)}}  \sum_{i=0}^T||CA_1^i B_1 - A_2^iB_2||^2_F = 
 \min_{\substack{C \in O(n)}}  ||CK_1 - K_2||^2_F
\end{equation}
We also provide a theoretical extension of Eq. \ref{eq: RedmanDSA} in Appendix \ref{app: Wasserstein}, which we note is highly susceptible to numerical instability. Although Eq. \ref{eq:DSA} requires iterative optimization, Eq. \ref{eq: \method} is solved via Procrustes alignment, which yields an \textit{exponential} acceleration of prior work. We provide further theoretical discussion in Appendix \ref{app: fastCcalc}. After solving for $C^*$, we can additionally study the joint state and input DSA scores: 
\begin{equation}\label{eq: stateDSA}
\text{\method}_{\text{state}}
(A_1, A_2,C^*) = ||C^*A_1C^{*^T} - A_2||_F^2
\end{equation}
\begin{equation}\label{eq: inputDSA}
\text{\method}_{\text{input}}
(B_1, B_2,C^*) = ||C^*B_1 - B_2||_F^2
\end{equation}

If the inputs directly applied to the system are known, as in computational models, Eq. \ref{eq: \method} is sufficient. However, when the true input is some modification of a surrogate input, it may be necessary to align the input as well. This is relevant in settings such as the comparison of two brain regions, when the surrogate input $u$ is a behavioral or sensory variable that is transformed by upstream regions. We therefore can extend Eq. \ref{eq: \method} to consider joint alignment of the input, without significant differences in the optimization problem. For further technical details, see Appendix \ref{app: misalignedinput}.

This metric motivates the following approach as in \cite{ostrow2024beyond}: identify the best linear approximation of an input-driven system, following which comparison can be done efficiently between the approximations. To do so, \cite{ostrow2024beyond} applied the Dynamic Mode Decomposition \citep{schmid2022dynamic}, which we introduce and extend to fit our setting next.


\subsection{Estimating Linear Operators}

As in DSA, We fit linear operators via the Dynamic Mode Decomposition (DMD) family of methods. The DMD \citep{schmid_dynamic_2010,schmid2022dynamic} identifies the linear dynamics that best explain the data:
\begin{equation}
    \phi(x_{t+1}) = A\phi(x_{t}).
\end{equation}
Here $x_t$ represents the measured state of the system at time $t$, $\phi$ is a nonlinear embedding of the data that typically expands the dimensionality of the state space, and $A$ is a matrix that is identified using some variant of least-squares regression. The goal of the Dynamic Mode Decomposition is to approximate the Koopman Operator \citep{Koopman1931}, a theoretical object that exists for all dynamical systems which encodes the linear dynamics of observables (functions that act on the state) under the system dynamics. 
Prior work has explored many different choices of $\phi$. For example, $\phi$ can be a kernel function, a delay embedding, or even a neural network \citep{williams_kernel-based_2016,brunton_chaos_2017,arbabi_ergodic_2017,lusch_deep_2018}. 
Intuitively, the dimensionality expansion acts similarly to the kernel trick \citep{smola1998learning}, where embedding into higher dimensions `unfolds' the nonlinearity. The DMD can be applied in non-autonomous systems, although this risks mixing driving and intrinsic dynamics \citep{proctor2016dynamic}. 

\paragraph{Incorporating Control into DMD and Koopman}
While the original Koopman theory assumed autonomous dynamics, prior work has sought to incorporate control into the theory \citep{mezickoopmancontrol1,proctor_generalizing_2016,strasser2025overviewkoopmanbasedcontrolerror,asada2024controlcoherentkoopmanmodelingphysical,haseli2025roadskoopmanoperatortheory}. Likewise, the DMD can be generalized to driven systems: 
When given control inputs $u_t$, we can instead apply DMD with control (DMDc, \citealt{proctor2016dynamic,proctor_generalizing_2016}):
\begin{equation}
    \phi_1(x_{t+1}) = A\phi_1(x_{t}) + B\phi_2(u_t)
\end{equation}
Here, $\phi_1$ and $\phi_2$ can be distinct nonlinearities. 
While DMDc was originally only applied with no nonlinearity ($\phi_1, \phi_2 = \text{Id}$), it too can be freely generalized to high-dimensional nonlinear embeddings. For more algorithmic details, see Appendix \ref{app: dmdc}. 

\paragraph{Issues of Partial Observation}
While estimating $A$ and $B$ via DMDc is an intuitive extension to input-driven systems, it has a hidden failure mode in the analysis of partially-observed systems. This is particularly important in the analysis of neural data, in which a small subset of neurons in a vast population are recorded. Generically, an input-driven system that is partially observed receives input to both the observed and unobserved components. The input at time $t$ therefore affects the observed state at time $t$ (instantaneously) and in future time steps through the unobserved state (Fig. \ref{fig:fig2}A). This means that simply applying DMDc in this setting will result in the $B$ matrix becoming biased toward the intrinsic dynamics of the system. We develop a formal description of this problem in Appendix \ref{app: partial obs bias}. We solve this problem by introducing \textbf{Subspace DMDc}, a novel extension of Subspace DMD \citep{takeishi_subspace_2017} that incorporates input. In brief, Subspace DMDc utilizes subspace identification algorithms from classical control theory \citep{Verhaegen_Verdult_2007}, which seek to identify linear dynamical systems of the form:
\begin{align}
    x_{t+1} &= Ax_t + Bu_t \quad \quad y_t = Cx_t
\end{align}

With only $y_t$ and $u_t$ observed. The situation of partial observability is a special case of this problem. In practice, we use the well-known N4SID or PO-MOESP algorithms to estimate $A$ and $B$ \citep{van1994n4sid,verhaegen1994identification} on  lifted states (thereby leveraging the power of nonlinear DMD algorithms such as \citealt{williams_kernel-based_2016}, although). For further technical details on the subspace identification algorithm, see Appendix Section \ref{app: subspace id}. We also highlight that subspace identification methods are designed to handle data with both observation and process noise, therefore providing \method noise robustness \citep{verhaegen1994identification,Verhaegen_Verdult_2007}. 


\section{Experiments}
\subsection{\method discriminates intrinsic dynamics from input-driven dynamics}

To demonstrate that \method can capture similarities in both intrinsic and input-driven dynamics, we simulated partially observed  nonlinear discrete-time systems with the following equations:
\begin{align}\label{eq: nonlin_po}
    x_{t+1} &= A(x_t + gF\tanh(x_t)) + B(u_t + \tanh(u_t)) \\
    y_t &= \begin{pmatrix} \mathbf{I}_{d} & \mathbf{0}_{n - d} \end{pmatrix}x_t + \epsilon_t
\end{align}

Where F and g are fixed across all simulations, and $\epsilon_t$ is observation noise. We randomly sampled two matrices for $A \in \mathbb{R}^{n \times n}$, and two for $B \in \mathbb{R}^{n \times 1}$, from which we constructed 4 systems: Systems 1 and 2 (3 and 4) share the same intrinsic dynamics matrix $A_1$ ($A_2$), while Systems 1 and 3 (2 and 4) share the same input matrix $B_1$ ($B_2$). We randomly sampled low-pass filtered white noise as the input drive (four times for each system), each with random initial conditions, yielding 16 systems in each distance matrix. In our experiments, we simulated 20-dimensional ($x \in \mathbb{R}^{20}$) systems and observed 2 dimensions ($y \in \mathbb{R}^2)$ for 5,000 time points. For simulation details, see Appendix \ref{app: po_comp}. We computed 5 distance matrices for each dataset, across 100 random seeds: (1), the DSA score using a delay-embedded DMD (Hankel DMD, or Hankel Alternative View of Koopman \citealt{arbabi_ergodic_2017,brunton_chaos_2017}), (2) the state distance using a delay-embedded DMDc, (3) the state distance using the SubspaceDMDc, (4) the input distance using the DMDc, and (5) the input distance using the SubspaceDMDc. Note that DSA does not have the ability to compare inputs, so it is left out. For a discussion on hyperparameter tuning, see Appendix \ref{app: po_comp}). For the sake of space, we report the jointly optimized input distance (Input DSA, Eq. \ref{eq: inputDSA}) and the individually optimized state distance (State DSA, Eq. \ref{eq: stateDSA}) as these are the most interpretable, although the jointly optimized state distance was highly similar. 

\begin{figure*}[!tbhp]
    \vspace{-10mm}
    \centering
    \includegraphics[width=\linewidth]{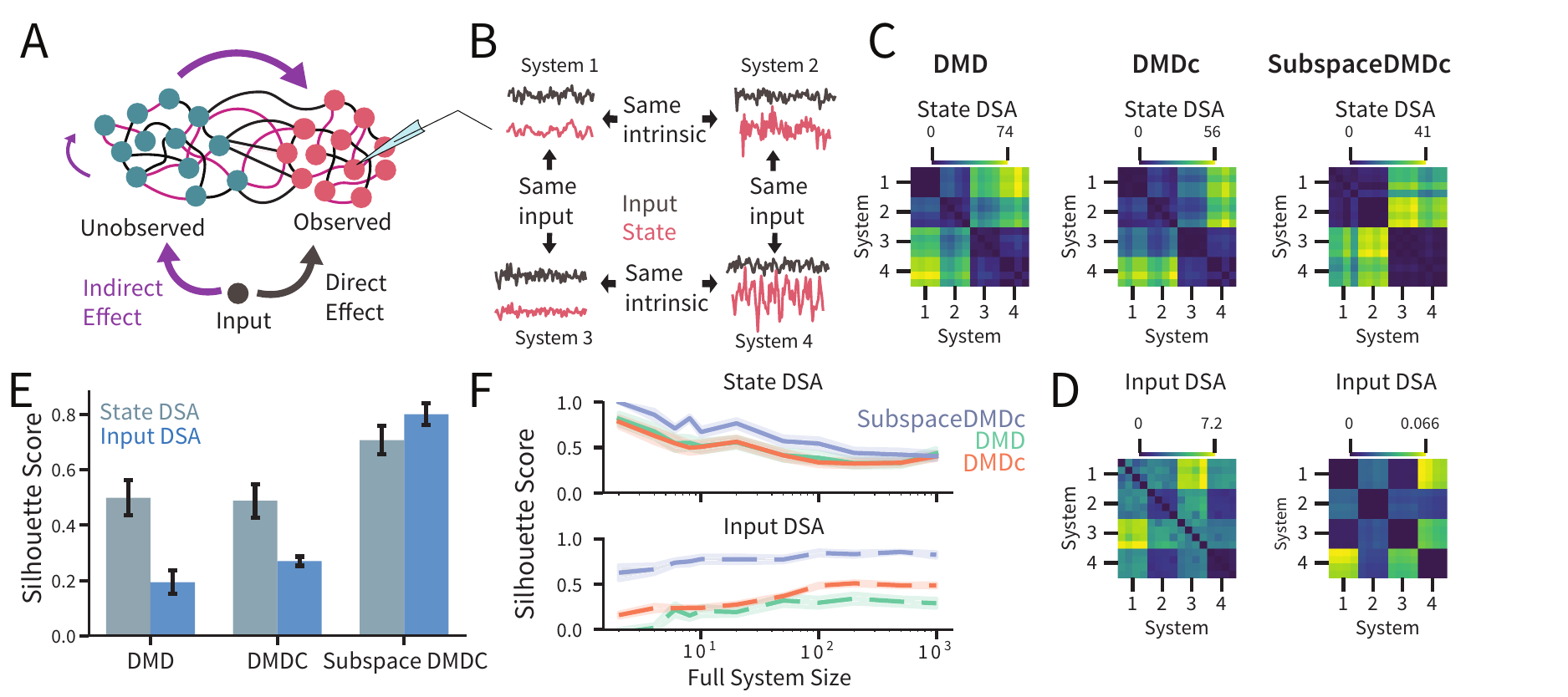}
    \caption{\textbf{\method SubspaceDMDc is robust to partial observation} \textbf{(A)} Under partial observation, inputs can have effects on observed states (red nodes) in the future via the unobserved states (green nodes), thereby biasing estimates of input driven-dynamics. Purple arrows indicate this indirect propagation of input into the observed states. \textbf{(B)} Sample inputs and observed states from 4 dynamical systems, which have alternate pairings of the same intrinsic and input-driven dynamics denoted by arrows. \textbf{(C)} Sample state similarity matrices based on estimates from the DMD, DMDc and SubspaceDMDc on data generated as in (B). four iterations of each system are generated, each with unique inputs and initial conditions, resulting in 16 x 16 dimensional matrices. \textbf{(D)} Sample input distance matrices on the same data as in (C). The DMD does not learn an input operator. \textbf{(E)} Aggregate silhouette scores of each similarity matrix across 100 random seeds, each generated as in (C,D). As a baseline input-label silhouette score for DMD is computed on the state matrix with the ground-truth input similarity labels. Bars denote standard error across seeds.  \textbf{(F)} Silhouette scores for each DMD and similarity type as the system is increased from 2-dimensional to 1000-dimensional. Each size was repeated across 20 random seeds. Shading denotes standard error. }
    \label{fig:fig2}
\end{figure*}

In Fig. \ref{fig:fig2}B we visualize the observed input and one dimension of the observed output for a sample set of systems, noting that it is not obvious at all a priori, let alone from the geometry, of any similarity relationships. We present sample state distance matrices from one random seed in Fig. \ref{fig:fig2}C. While the DMD and the DMDc have notable structure pertaining to the true state similarity, the SubspaceDMDc similarity scores are noticeably sharper. Quantifying these matrices with the silhouette score (a measure of cluster separability and dispersal, 1.0 is best) utilizing ground-truth state labels, the DMD scores 0.6, DMDc scores 0.68, and the Subspace DMDc scores 0.94. In Fig. \ref{fig:fig2}D, we present the respective input scores for each method. As predicted by our previous discussion on the effects of partial observation on input matrix estimation, the input DSA score computed with DMDc does not align with ground truth, reporting a silhouette score of 0.19. The silhouette score of the SubspaceDMDc is 0.83, indicating robust separation. We also compute the total similarity matrices (Eq. \ref{eq: \method}, Appendix Fig. \ref{app: fig: jointdsa}), for which the SubspaceDMDc reports correctly that each type of system is altogether unique. We swept over 100 seeds in Fig. \ref{fig:fig2}E and found that the SubspaceDMDc-based \method consistently yielded the best separability. 

To assess the effect of partial observation, we ran the above analysis for different-sized systems (ranging from 2 to 1000 dimensions) with only 2 observed dimensions, for which we present the average silhouette scores for \method in Fig. \ref{fig:fig2}F. The state similarity scores for each method gracefully degrade with the total state size, and SubspaceDMDc has a noticeable improvement over the other methods. The DMDc input score appears to never be robust. However, the SubspaceDMDc input similarity is robust across all system sizes. This suggests that SubspaceDMDc can be used to measure the dynamical similarity of partially observed, input-driven dynamical systems. 

\subsection{Robustness to input noise and transformation}

In real-world settings such as neural populations, the true inputs driving the system are rarely accessible. Instead, what we observe are often noisy or partial measurements, limited by sensor resolution, sampling rates, or inherent partial observability. As a consequence, researchers often rely on behavioral variables, task instructions, or environmental features as proxies when modeling neural circuits \citep{vinograd2024causal, sani2024dissociative,burak_accurate_2009,schaeffer_reverse-engineering_2020,mante_context-dependent_2013}. This raises a key question for applying \method: if the true inputs are unknown, can \textit{surrogate inputs} that are correlated with the ground truth still yield accurate distance estimates?

\begin{figure*}[tbhp!]
    \centering
    \vspace{-10mm}
    \includegraphics[width=\linewidth]{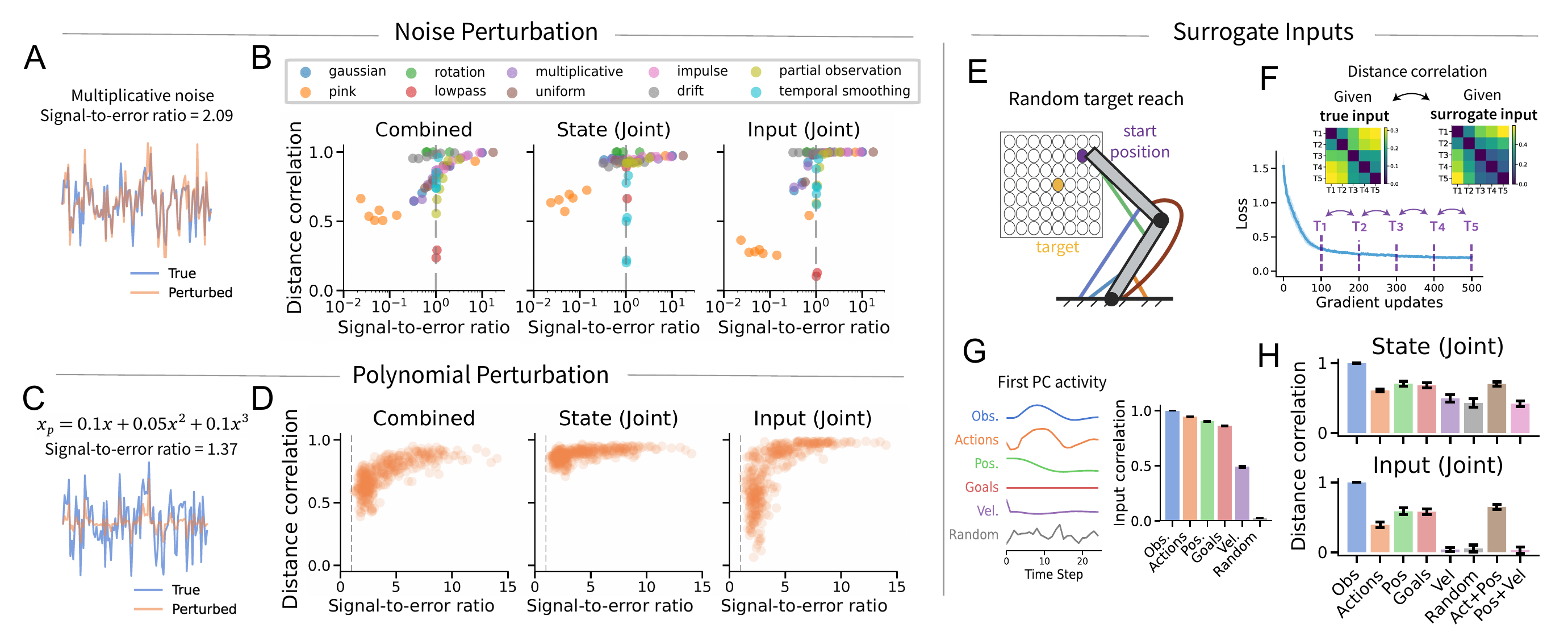}
    \vspace{-5mm}
    \caption{\textbf{\method provides robust distance estimates under input noise and surrogate inputs.} 
    \textbf{(A)} Example of multiplicative Gaussian noise added to input data.  
    \textbf{(B)} Effect of different noise perturbations on the \method similarity matrices in Fig. \ref{fig:fig2} (see Appendix Section \ref{app:input_noise} for further technical details on the noise). The y-axis indicates the correlation between the \method matrices given the true input and the perturbed input. The x-axis indicates the signal to error ratio $\text{Var(X)} / \text{Var}(\tilde{X} - X)$. From left to right: joint controllability DSA (Eq. \ref{eq: \method}), jointly optimized state DSA (Eq. \ref{eq: stateDSA}; jointly optimized input DSA (Eq. \ref{eq: inputDSA}). 
    \textbf{(C)} Example of a polynomial function applied to the same input as in (A). 
    \textbf{(D)} Similar analysis as in (B), with various random polynomial functions applied to the input. 
    \textbf{(E)} Random target task schematic.
    \textbf{(F)} We compare RNNs dynamics across multiple time points in training with \method. We study changes in the distance matrix when applying surrogate inputs.
    \textbf{(G)} Example first Principal Component for different surrogate inputs and their correlation with the true input (Obs).
    \textbf{(H}) Correlation between \method distances estimated using the ground truth input and surrogate inputs. Error bars indicate standard error across 10 training runs. Jointly optimized state and input DSA are presented.
}
    \label{fig:fig3}
    \vspace{-5mm}
\end{figure*}

We begin by examining how well the true \method distance matrix when the provided input is noise-corrupted. We repeated the simulation and comparison in Fig. \ref{fig:fig2}C, this time applying different types of noise perturbations to the input used in SubspaceDMDc (example in \ref{fig:fig3}A). For complete details on the noise perturbations, see Appendix \ref{app:input_noise}.  We applied 10 types of perturbations inspired by different real-world situations, such as partial observation, temporal smoothing, or multiplicative Gaussian noise, and repeated each perturbation across a range of parameters (e.g. standard deviation in the noise settings or filter width in the smoothing setting). 
To measure the deviation of the signal consistently across perturbation types, we compute the signal-to-error ratio (SER) for each perturbation: given a time series $X \in \mathbb{R}^{t \times d}$ and its perturbed version $\tilde{X} = f(X)$, SER is defined as 
$$SER(X,\tilde{X}) = \frac{\text{Var}(X)}{\text{Var}(\tilde{X} - X)}$$ 
SER generalizes signal-to-noise ratio for non-additive perturbations. Despite the prevalence of noisy inputs, we found that \method distances remain robust, decaying slowly below the $SER < 1$ threshold (Fig. \ref{fig:fig3}B): High SERs lead to high correlations with ground truth distances, and correlations tend remain above $r > 0.75$ even as SER approaches 1. This robustness arises due to the delay embedding and reduced-rank regression in Subspace DMDc: delay embedding incorporates the history of inputs, while reduced-rank regression removes noisy modes with spurious correlations. To generalize this analysis to more complicated transformations, we repeated the analysis using inputs transformed by random polynomials (Fig. \ref{fig:fig3}C,D). Specifically, we sampled 500 random 4-th order polynomials with coefficients drawn uniformly from [$-0.1, 0.1$], which we applied dimension-wise to the inputs as a new perturbation. To generate inputs with higher SERs, we also generated 200 polynomials where the linear coefficient was fixed at $0.9$, while all other coefficients were sampled from the same range.
As in the previous analysis, we find a similar pattern across SER: the state DSA correlations are the most robust, followed by the combined and the input DSA scores. Together, this suggests that up to reasonable expectations (SER greater than or close to 1), the \method scores are robust to generic perturbations on the input data.



Next, we evaluated whether task-relevant surrogate inputs could be used in place of ground truth, instead of perturbed versions of the true input. We analyzed trained RNNs from the Random Target Reach task (Fig. \ref{fig:fig3}E, \citealt{Codol2024MotorNet}), a widely used paradigm for studying neural control of movement from which rich neural and behavioral dynamics emerge \citep{Hatsopoulos2007Fragments, Flint2012FieldPotentials, Churchland2012Nature}. Across 20 epochs equally spaced in training, we recorded the RNN’s hidden states, observations (the true input), actions (behavioral output), and other task variables (Fig. \ref{fig:fig3}F, only 5 epochs shown for visualization purpose). For a detailed description of the task and training, see Appendix \ref{app:random_target}. Passing the hidden states of the RNN and the ground truth inputs through \method, we obtained two distance matrices that characterize how the network's intrinsic and input-driven dynamics change over learning. We repeated this process for various task-related surrogate inputs: RNN output (actions), position, velocity, task instruction, and various combinations. We also included random inputs sampled from the uniform distribution on $[0,1]$ as a baseline. Among surrogates, the actions maintain the highest trial-averaged correlation with the ground truth input (Fig. \ref{fig:fig3}G). We find that \method intrinsic (state) distances estimated with surrogate inputs have strong correlation with the ground truth distance, even with random inputs (Fig. \ref{fig:fig3}H). For input-driven comparisons, more highly-correlated surrogates tend to yield more accurate similarities, with the RNN’s combined action and positions providing strong correlations with the ground truth distance (Fig. \ref{fig:fig3}H). Overall, our robustness analysis suggests that state similarities are highly robust to perturbations of many different types, while the combined and input similarities are still robust, albeit less so. 

\section{Applications}
\vspace{-5pt}

\subsection{\method { }tracks the evolution of individual difference over learning}

\begin{wrapfigure}{r}{0.65\textwidth}
    \centering
    \includegraphics[width=\linewidth]{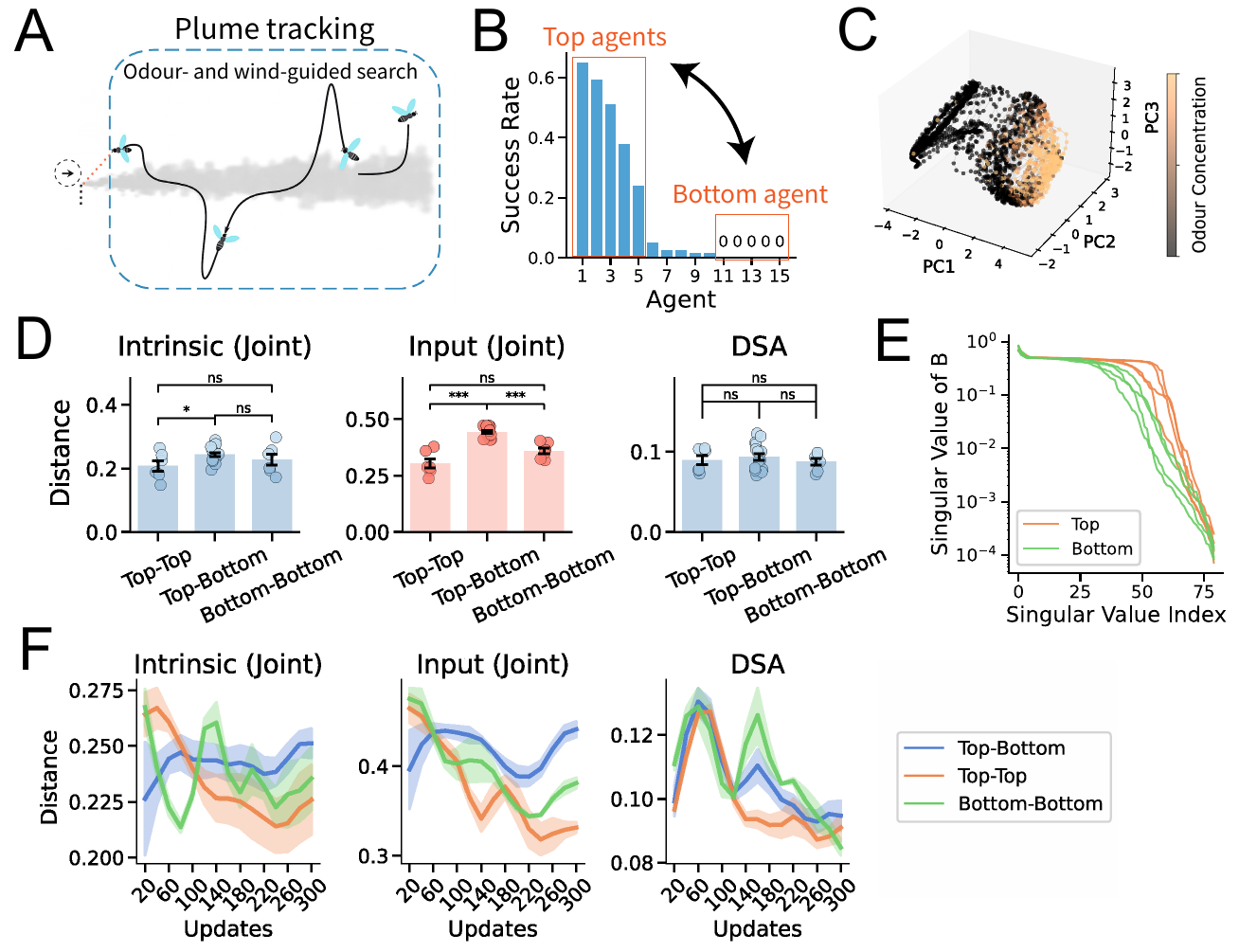}
    \caption{\textbf{\method identifies how successful and unsuccessful agents differ over training. (A)} The Plume Tracking environment schematic adapted from \cite{singh2023emergent}. 
    \textbf{(B)} Average performance (success rate) of 15 independently trained agents. The 5 most performant ("Top") and 5 failed ("Bottom") agents are studied further.
   \textbf{ (C)} Neural dynamics of trained agents are organized in a low-dimensional space and reflective of behaviorally relevant variable (i.e. the odor concentration). 
\textbf{   (D)} Average distance computed within the 5 Top Agents, within the 5 Bottom agents, and across groups (Top–Bottom). 
\textbf{(E)}  The singular value spectrum of the input-mapping operator $B$ from Top and Bottom agents. 
\textbf{(F)} The evolution of similarity within and across groups over learning. Shaded area indicates standard error.}
    \vspace{-10mm}
    \label{fig:fig5}
\end{wrapfigure}

In closed-loop Reinforcement Learning (RL) environments, stochastic action selection and small differences in policies can shift the distribution of sensory inputs encountered across training. To understand divergence between agents, it is crucial to
how inputs interact with network dynamics and shape agent performance. The Plume Tracking task (Fig. \ref{fig:fig5}A) provides an ideal testbed because the agents must balance between memory-based intrinsic dynamics with stimulus-driven responses. 

In this task, artificial flies (RNNs) trained by deep RL navigate to the source of a simulated turbulent odor plume in a windy 2D arena. At each timestep, the agent senses only local cues (intermittent odor concentration and wind direction) and takes actions to move its position. Due to the stochastic nature of sensory observations and exploration, agents diverge across training, producing a wide variation of success rates (Fig. \ref{fig:fig3}B). This raises a key question: do performance differences reflect variations in intrinsic dynamics (the ability to form and maintain task-relevant representations) or input-driven responses to stimuli?

We trained 15 independent agents on the Plume Tracking task. We selected the five best-performing (“Top”) agents with 65\% to 20\% success rate at locating the odor source across 200 evaluation episodes, and five worst-performing (“Bottom”) agents who never succeeded on any episode (Fig.  \ref{fig:fig5}B). Applying \method revealed that the input-driven dynamics of the Top agents were significantly more similar to each other and clearly separated from those of the Bottom agents, whereas intrinsic dynamics were not significantly different between groups (Fig.  \ref{fig:fig5}D). This suggests that successful plume tracking heavily depends rapid responses to wind direction and odor concentration. To probe how the input-driven dynamics differ between Top and Bottom agents, we examined the singular values of the input–mapping \(B\) in Fig. \ref{fig:fig5}E. Singular values of the operator quantify how strongly input directions are injected into RNN state space. We found that the singular values of \(B\) for Top agents decay more slowly than for Bottom agents. This implies that inputs excite more dimensions of the RNN in Top agents (Fig.  \ref{fig:fig5}E), thereby allowing them to more effectively incorporate recent information to inform action selection in the future. 

We next ask how individual variability in neural dynamics evolves during training. To this end, we computed pairwise dynamical similarity among Top and Bottom agents every 20 gradient updates (Fig.  \ref{fig:fig5}F). While within–group input similarity decreases over training for both Top and Bottom, the Top agents ultimately converge to a more consistent set of input-driven dynamics, whereas the Bottom agents diverge toward heterogeneous, idiosyncratic dynamics. This is reminiscent of the "Anna Karenina principle", in which effective solutions are similar to each other, while worse ones are highly varied.

\subsection{\method captures differences in neural population dynamics across time}

\begin{figure*}[h] 
    \centering
    \includegraphics[width=\linewidth]{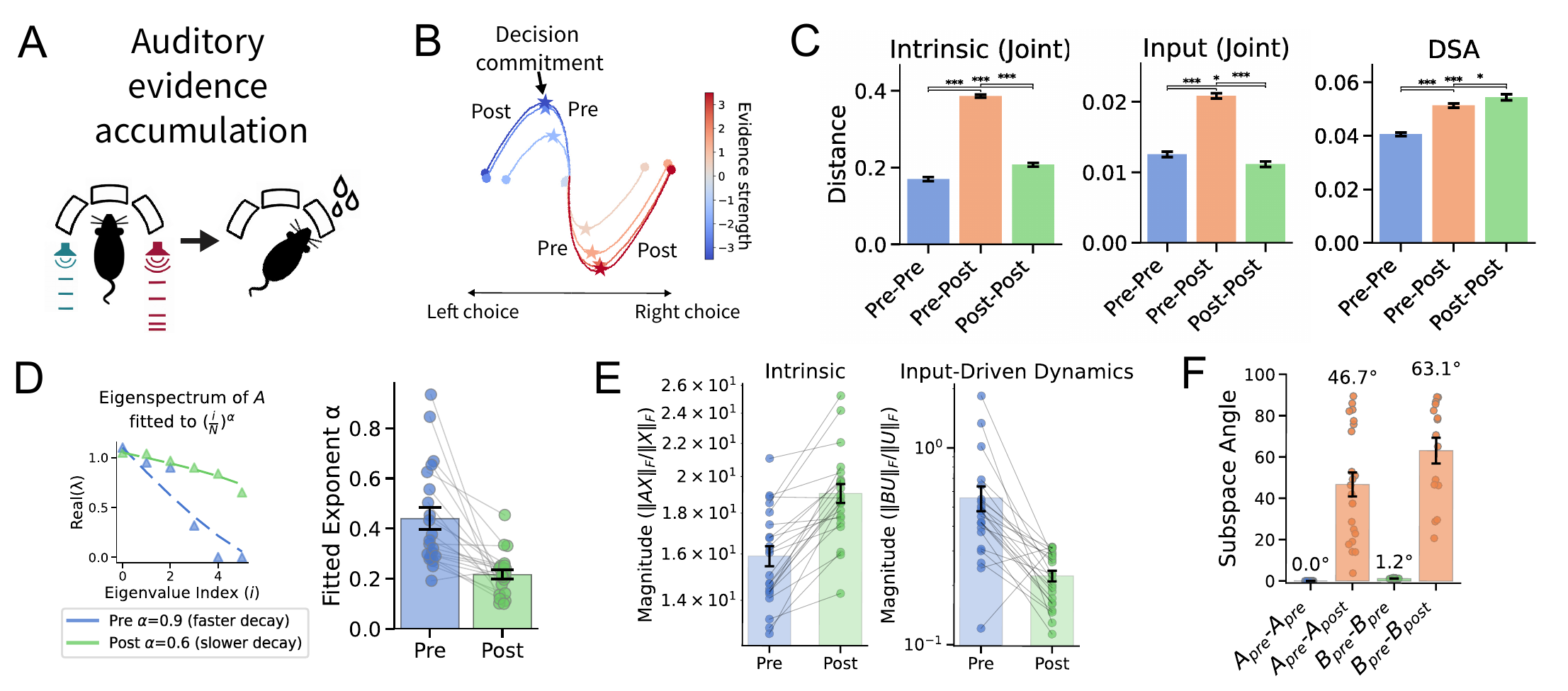}
    \caption{\textbf{\method quantifies differences in neural population dynamics across task epoch.} 
    \textbf{(A)} Auditory evidence accumulation task schematic (adapted from \citealt{luo2025transitions}). 
   \textbf{ (B)} Trial-averaged neural trajectories visualized in the top two Principal Components. Stars indicate a "neural time of commitment" (\textit{nTc}): the time point when the curvature of trial-averaged trajectories is maximum (marked by stars).
    \textbf{(C)} Similarity of neural dynamics before and after the nTc for rat T223. Bars denote standard error across 21 sessions. 
    \textbf{(D)} Distribution of top real eigenvalues of state-transition matrix $A$ and fit power law for pre- vs. post-commitment activity. Left, sample distribution. Right, distribution of power law exponents across sessions. Dots denote individual sessions, lines indicate paired periods within session, likewise in E and F. 
    \textbf{(E)} Normalized effects of intrinsic and input-driven dynamics in pre vs. post periods. 
    \textbf{(F)} Subspace angles of the input and state operators within and between time periods. $A_{pre}-A_{pre}$ (likewise $B_{pre}$) denotes the noise floor via split-halves comparison.
    }
    \vspace{-4mm}
    \label{fig:fig6}
\end{figure*}

Lastly, we apply \method to a recently published dataset in which neural population activities were recorded from six frontal and striatal regions with Neuropixels probes during an auditory evidence accumulation task (\citealt{luo2025transitions}, Fig.  \ref{fig:fig6}A). During this task, rats were trained to listen to auditory pulses from speakers on the either side of the animal, and to turn to the side with more auditory pulses. This dataset contains 12 rats across 115 daily sessions with a median of 327 neurons recorded and 455 trials completed per session. We chose 4 rats with more than 20 recorded sessions for our analysis to ensure accurate estimation of neural dynamics. In the original study, the authors define the \textit{neural time of commitment (nTc)}  as the internal moment during perceptual decision-making when an animal has effectively committed to a choice (Fig.  \ref{fig:fig6}B). To examine how neural population dynamics reorganize across this point, we applied \method to neural activities aligned to the \textit{nTc}. Spiking activity was binned in 50 ms windows, smoothed with a causal Gaussian kernel ($\sigma$ = 250 ms), and dimensionality reduced with PCA to preserve 99\% of variance. The activity was then embedded into three dimensions using Isomap, and \method was applied with hyperparameters detailed in Appendix \ref{app:hyperparam_Luo}. We construct the inputs as two-dimensional time series encoding the number of auditory pulses from the left and right within each bin.

Comparing neural dynamics before and after the \textit{nTc} (“Pre” vs. “Post”), we found significant shifts in both intrinsic and input-driven dynamics, consistent with the changes at \textit{nTc} reported in \citep{luo2025transitions} (Fig.  \ref{fig:fig6}C). To probe how the intrinsic dynamics change, we analyzed the eigenspectrum of the state-transition matrix $A$ estimated by SubspaceDMDc before and after the \textit{nTc}. Each spectrum was fit with a power law \(\lambda_i \;\propto\; \left(\frac{i}{N}\right)^{\alpha}, 
\) where $\lambda_i$ are the $i$-th eigenvalue sorted in descending order. We found that the post-commitment periods consistently showed smaller $\alpha$, indicating slower decay and thus longer-lasting intrinsic dynamics (Fig.  \ref{fig:fig6}D). This is directly related to the \textit{controllability} of the dynamical system, which describes how easy it is for an input sequence to drive the system to arbitrary points in state space \citep{luenberger1979dynamic}. Smaller DMD eigenvalues implies greater input controllability, which would be expected for a more input-driven system as \citet{luo2025transitions} identified is the case in the pre-\textit{nTc} regime. 

Likewise, the average magnitude of the intrinsic dynamics strengthen while the input-driven dynamics weaken in the Post-nTC period, reflecting a transition into more autonomous, less input-sensitive regime after the \textit{nTc} (Fig.  \ref{fig:fig6}E). 
By measuring the subspace angle between the $A$ and $B$ operators across the \textit{nTc}, we find that the neural dynamics occupy different subspaces across the \textit{nTc}, reflected a reorganization of dynamics at the decision-commitment time (Fig. \ref{fig:fig6}F). Together, these results suggest that population activity undergoes a regime shift at the \textit{nTc}: transitioning from an input-driven, evidence-accumulation phase into an intrinsically dominated, decision-commitment phase, as suggested by \citep{luo2025transitions}.


\vspace{-2mm}
\section{Discussion}
\vspace{-2mm}
We introduced a theoretically-motivated method (\method) to quantitatively compare the intrinsic dynamics and effects of input between two dynamical systems, from data alone. We extended the DSA framework \citep{ostrow2024beyond} to account for input-driven systems, which required a novel variant of the Dynamic Mode Decomposition with Control \citep{proctor2016dynamic} called Subspace DMDc. We also developed a novel optimization algorithm for our similarity metric that is multiple orders of magnitude faster than prior work.    

We demonstrated that \method can effectively estimate similarity from partially-observed systems (Fig. \ref{fig:fig2}), which is necessary when dealing with most physical and biological systems. In many settings, the true input is not known (for example the signal from one brain region to another), but we demonstrated that even approximate or noisy inputs can provide reasonable input and intrinsic similarity estimates (Fig. \ref{fig:fig3}). Since many models in computational neuroscience tend to utilize proxy inputs \citep{nair2023approximate,sohn2019bayesian,burak2009accurate,mante2013context,sussillo2015neural}, our work provides principled methodological support to this practice. Inputs could also be estimated via another computational method (e.g. \citealt{perich2020inferring, luo2025transitions}) before applying \method. As Fig. \ref{fig:fig3} suggests, even utilizing weakly correlated proxy inputs can increase the robustness of the intrinsic comparison with \method. 


\method could be used for further validation of computational models with perturbation as in \cite{o2022direct}. Known optogenetic or electrical impulse perturbations could be applied to both a model and biological neural circuit, following which both their internal dynamics and impulse responses could be compared. This can provide more stringent tests than comparing intrinsic dynamics alone. Other subspace identification methods could be used in place of SubspaceDMDc, such as Eigensystem Realization (ERA, \citealt{juang_eigensystem_1985}). In a similar vein, \method could potentially be used to identify the information content in cross-brain-region communication -- multiple models could be constructed with different surrogate inputs, and the most similar input should have the lowest input distance to the data (Fig. \ref{fig:fig3}). 

Although we only applied \method to biological neural data and recurrent neural networks, it can be applied to any time series data. Indeed, the constraints on the method are based on the capabilities of systems identification and Koopman Operator approximation. For example, if the input is not persistently exciting, state modes will be under-approximated. If a viable basis is not identified, the linear model may not be able to capture enough structure for effective comparison. However, there exists a wide range of work in both fields designed to tackle these problems \citep{dmdchallenges,Colbrook_2023,takeishi_learning_2017,ichinaga_pydmd_2024}. It is also worth noting that near-perfect estimation is not necessary for informative comparison. 

\method has other limitations. The method assumes additive input, which may not be able to approximate the effects of multiplicative input \citep{logiaco2021thalamic,shine2021computational}. Disentangling the contribution of state and input can also be challenging or intractable when they are synchronized \citep{RajanEtAl2010} or the input is a linear function of the state (\citealt{Verhaegen_Verdult_2007}, although methods exist for subspace identification in closed loop \citealt{van_der_veen_closedloop_2013}).  From a computational complexity standpoint, the bottleneck is fitting the SubspaceDMDc, as comparison is extremely fast. Regardless, we found that even for reasonably sized systems (e.g. 50 dimensions, 10,000 timepoints) and hyperparameters (100 delays), the method requires a O(1 minute) on M1 Pro Mac, and is even faster on a GPU.

\clearpage

\section*{Acknowledgements}
We thank members of the Rajan and Fiete labs for helpful discussions.
Funded by NIH (RF1DA056403 to K.R.), 
James S. McDonnell Foundation (220020466 to K.R.), 
Simons Foundation (Pilot Extension-00003332-02 to K.R.), 
McKnight Endowment Fund (K.R.), 
CIFAR Azrieli Global Scholar Program (K.R.), 
NSF (2046583 to K.R.), 
Harvard Medical School Neurobiology Lefler Small Grant Award (K.R.),
Harvard Medical School Dean's Innovation Award (K.R. and S.H.S.), 
and Alice and Joseph Brooks Fund Postdoctoral Fellowship (S.H.S.). 
A.H is supported by the Kempner Graduate Fellowship.
M.O. is funded by the NSF GRFP.


\bibliography{main}
\bibliographystyle{iclr2026_conference}

\clearpage

\appendix
\include{appendix}

\end{document}

%% file: math_commands.tex
\usepackage{amsmath,amsfonts,bm}

\def\eqref#1{equation~\ref{#1}}

\def\1{\bm{1}}

\DeclareMathAlphabet{\mathsfit}{\encodingdefault}{\sfdefault}{m}{sl}
\SetMathAlphabet{\mathsfit}{bold}{\encodingdefault}{\sfdefault}{bx}{n}



%% file: appendix.tex
{\Large \centering Appendix \\}

\section{LLM Usage Statement}
We used LLMs in preliminary phases of conducting this research, in particular for brainstorming research ideas and literature review, as well as writing simple boilerplate code (e.g. plotting). All code, math, and writing was checked by at least one author before including it in the paper.
\section{Dynamic Mode Decomposition with Control (DMDc)} \label{app: dmdc}

Dynamic Mode Decomposition with control (DMDc) \citep{proctor2016dynamic} extends
standard DMD to dynamical systems with external inputs. It provides a data-driven approximation of both the intrinsic dynamics $A$ and input couplings $B$, enabling system identification and forecasting for non-autonomous dynamical systems. Here, we briefly review the formulation of DMDc. For full details, please refer to \cite{proctor2016dynamic}. In practice, we can apply DMDc whenever the state is fully observed. When this is not the case, refer to Sections \ref{app: partial obs bias} and \ref{app: subspace id}.

We consider the input-driven linear model
\begin{align}
\label{eq: dmdc}
x_{k+1} = A x_k + B u_k, \qquad 
A \in \mathbb{R}^{n \times n}, \;\; B \in \mathbb{R}^{n \times p},
\end{align} 
where $x_k \in \mathbb{R}^n$ are state snapshots and $u_k \in \mathbb{R}^p$ are input signals.  For an input-driven dynamical system, we collect pairs of the system states and input signals into
\begin{align}
X &= \begin{bmatrix} x_1 & x_2 & \cdots & x_{m-1} \end{bmatrix}, \\
X' &= \begin{bmatrix} x_2 & x_3 & \cdots & x_m \end{bmatrix}, \\
U &= \begin{bmatrix} u_1 & u_2 & \cdots & u_{m-1} \end{bmatrix},
\end{align}
where $X, X' \in \mathbb{R}^{n \times (m-1)}$ and $U \in \mathbb{R}^{p \times (m-1)}$. We can rewrite equation \ref{eq: dmdc} into 
\begin{align}
X' = G \Omega = \begin{bmatrix} A & B \end{bmatrix} \begin{bmatrix} X \\ U \end{bmatrix}
\end{align}
where $\Omega \in \mathbb{R}^{(n+p) \times (m-1)}$ and $G \in \mathbb{R}^{n \times (n+p)} $. 

The optimal operator is then obtained by solving
\begin{align}
G = \arg\min_{\tilde G} \|X' - \tilde G \Omega\|_F
= X' \, \Omega^{+},
\end{align}
where $(\cdot)^{+}$ denotes the Moore--Penrose pseudoinverse.

Let the truncated SVD of $\Omega$ be
\begin{align}
\Omega \approx \tilde U \, \tilde \Sigma \, \tilde V^*,
\qquad
\Omega^{+} \approx \tilde V \, \tilde \Sigma^{-1} \tilde U^*.
\end{align}
Partition $\tilde U$ into state and input blocks:
\begin{align}
\tilde U =
\begin{bmatrix}
\tilde U_x \\[2pt] \tilde U_u
\end{bmatrix}, \qquad
\tilde U_x \in \mathbb{R}^{n \times \tilde r}, \;
\tilde U_u \in \mathbb{R}^{p \times \tilde r}.
\end{align}
The system matrices are then estimated as
\begin{align}
A = X' \tilde V \tilde \Sigma^{-1} \tilde U_x^*, \qquad
B = X' \tilde V \tilde \Sigma^{-1} \tilde U_u^*.
\end{align}

We can further project $A$ and $B$ into the system's state space using 
\begin{align}
X \approx U_r \Sigma_r V_r^*, \qquad U_r \in \mathbb{R}^{n \times r},
\end{align}
\begin{align}
\tilde A = U_r^* A U_r, \qquad
\tilde B = U_r^* B.
\end{align}

It is useful to perform SVD independently on $X$ and $U$, assuming there is minimal correlation among the variables. This is especially useful when using nonlinear embeddings such as delay embeddings in the regression. This changes the algorithm of DMDc but not significantly. In particular, we can now write:

\begin{align}
    \Omega = \begin{pmatrix} U_x & 0 \\ 0 & U_u \end{pmatrix} \begin{pmatrix} \Sigma_x & 0 \\ 0 & \Sigma_u \end{pmatrix} \begin{pmatrix} V_x^T  \\ V_u^T \end{pmatrix}
\end{align}

This enables us to pick ranks separately for $X$ and $U$ components. In practice, we apply the techniques used in HAVOK \citep{brunton_chaos_2017} to estimate the DMD. We do regression in the eigen-time-delay (pca-whitened) spaces of X and U (Hankelized), which allows us to select ranks separately for the X and the U space. 

\subsection{On nonlinear embeddings in DMDc} \label{app: input embeddings}

In the standard DMDc formulation (above), an SVD is taken across $\Omega$, which concatenates the state data $X$ with the input data $U$. Although this has the benefit of whitening across all regressors, it can bias the estimation of $A$ and $B$ depending on the relative scalings and dimensionalities of $X$ and $U$. This has a critical effect when applying high-dimensional nonlinear embeddings to only $X$ ($U$) individually, as the SVD will be increasingly dominated by signal from $X$ ($U$) if the data is sufficiently rich. Therefore, whenever we apply delay embeddings or other nonlinear embeddings to $X$, we do so commensurately to $U$.

\section{Relationship between DMD (regular) and DMDc}

\begin{equation*}
A_x^c = \left[ \begin{pmatrix} X^T X & X^T U \\ U^T X & U^T U \end{pmatrix}^{-1}_{1:m} \begin{pmatrix} X^T \\ U^T \end{pmatrix} \right] X_{n+1} 
\end{equation*}

\begin{equation*}
A_x = (X^T X)^{-1} X^T X_{n+1}
\end{equation*}

\begin{equation*}
S = \left(U^T U - U^T X (X^T X)^{-1} X^T U\right)
\end{equation*}

\begin{equation*}
A_x^c = \left[ \left( (X^T X)^{-1} + \overbrace{(X^T X)^{-1} X^T U}^{A_u} S^{-1} U^T X (X^T X)^{-1}, -(X^T X)^{-1} X^T U S^{-1} \right) \begin{pmatrix} X^T \\ U^T \end{pmatrix} \right] X_{n+1}
\end{equation*}

\begin{equation*}
A_x^c = (X^T X)^{-1} X^T X_{n+1} + A_u S^{-1} A_u^T X^T X_{n+1} - A_uS^{-1}U^TX_{n+1}
\end{equation*}

\begin{equation*}
A_x^c = A_x + A_u S^{-1} A_u^T X^T X_{n+1} - A_uS^{-1}U^TX_{n+1}
\end{equation*}

\section{Partial Observation Induces Biases in Input Operator B} \label{app: partial obs bias}

Consider a partially observed linear system:

\begin{align}
    \begin{pmatrix} x^{o} \\ x^{u} \end{pmatrix}_t = \begin{pmatrix} A_{oo} & A_{ou} \\ A_{uo} & A_{uu} \end{pmatrix} \begin{pmatrix} x^{o} \\ x^{u} \end{pmatrix}_{t-1} + \begin{pmatrix} B_{o} \\ B_u \end{pmatrix} u_{t-1}
\end{align}

We observe states $x^o$. This system can also be formulated as a Vector-Autoregressive model with exogenous inputs (VAR-X). To see this formulation, we recursively substitute the definition of $x^u_t$ with its dynamical equation, hence arriving at a formulation of $x^o_t$ as a function of past observed states and inputs:

\begin{align}
x^o_t &= A_o x^o_{t-1} + A_{ou}x^u_t + B_o u_t \\
    &= A_o x^o_{t-1} + A_{ou}[A_{uo}x^o_{t-2} + A_{uu}x^u_{t-2} + B_u u_{t-1}] \\
    &= \dots \\
    &= A_o x^o_{t-1} + \sum_{i=1}^\infty A_{ou}A_u^{i-1}(A_{uo}x^o_{t-1} + B_u u_{t-i}) + B_ou_t
\end{align}

We take an infinite sum here for completeness, but in practice $i$ can be capped up to marginal error based on the decay rates (eigenvalues) of $A_u$. We can write this equation as a function of the delay-embedded observed state and inputs:

\begin{align}
    x^o_{t+1} = \begin{bmatrix} A_o & A_{ou}A_{uo} & A_{ou} A_u A_{uo} & \dots & A_{ou}A_u^{d-1}A_{uo} \end{bmatrix} \begin{bmatrix} x^o_t \\ x_{t-1}^o \\ \dots \\ x_{t-d}^o \end{bmatrix} \\ 
    + \begin{bmatrix} B_o & A_{ou}B_u &  A_{ou}A_u B_u \dots A_{ou}A_u^{d-1} B_u \end{bmatrix} \begin{bmatrix} u_t \\ u_{t-1} \\ \dots \\ u_{t-d} \end{bmatrix} 
\end{align}

These equations show that when performing regression as in $DMDc$ on partially-observed, delay-embedded data, the estimates of $B$ become biased by the intrinsic dynamics in the unobserved states. Biases in $B$ emerges when utilizing delay embeddings as dimensionality expansions, as we can see from the above formulation. Although we display the formal connection with linear systems above, it is simple to observe that the same problem occurs with nonlinear dynamics as well. 

\section{De-Biasing $B$ under partial observation with Subspace Identification} \label{app: subspace id}

In this section, we introduce SubspaceDMDc, a natural extension of two DMD models in the literature: Subspace DMD \citep{takeishi_subspace_2017} and DMDc \citep{proctor2016dynamic}. SubspaceDMDc has a notable difference from SubspaceDMD, as \citet{takeishi_subspace_2017} utilize the subspace identification approach to handle observation noise, whereas we utilize subspace identification to handle input affecting future timesteps (although we gain noise robustness through similar means). In the control theory literature, there are a number of subspace identification algorithms, two of the most famous are Multivariable Output-Error State sPace (MOESP) modeling and Numerical Algorithms for Subspace State Space System Identification (N4SID) \citep{Verhaegen_Verdult_2007,verhaegen1994identification,van1994n4sid}. In order to be brief, we will discuss only N4SID, which is the method we chose to implement. In general, the algorithms have similar behavior, except on ill-conditioned data. Practically speaking, either method could be used in DSA; it is up to the user and their respective performances on the dataset. The extension of these methods to SubspaceDMDc is the introduction of a lifting feature space: polynomials, kernels, random feature maps, neural networks, or nonlinear features can be used in order to find a best-predicting nonlinear basis upon which the features evolve linearly. 

\subsection{Subspace DMD}
Subspace DMD \citep{takeishi_subspace_2017} is designed to handle the estimation of the Koopman operator given data that is contaminated with observational and process noise. Assuming that the dynamics and the noise are independent, one can project out the contribution of the noise in the data and leave only the component that is explainable with past data (via delay embedding, step 2 of the algorithm below). We assume real data, although the method works for complex data as well. Algorithm 2 in the paper reads:
\begin{enumerate}
    \item Construct data matrices $Y_p = \begin{bmatrix}Y_0^T & Y_1^T \end{bmatrix}^T \quad Y_f = \begin{bmatrix}Y_2^T & Y_3^T \end{bmatrix}^T $ where $Y_t = [g(x_t) \dots g(x_{t-m+1})]$
    \item Compute the orthogonal projection of the future data onto the past data: $O = Y_f \mathbb{P}_{Y^T_p}$ where the projector $\mathbb{P}_{Y^T_p} = Y_p^T(Y_pY_p^T)^\dagger Y_p$.
    \item Compute the compact SVD (e.g., the SVD with no zero rows or columns): $O = U_qS_qV_q^T$ and define $U_{q1}, U_{q2}$ by taking the first and last $n$ rows of $U_q$. This is done in order to split the projection matrix into the observability matrix and the state matrix: $O = \Gamma X$, up to right / left multiplication by an invertible matrix. The observability matrix looks like $\Gamma = \begin{pmatrix} C & CA & \dots & CA^n \end{pmatrix}$. Because this matrix encodes the time-shifted structure of the dynamics, we split into the top n and last n rows to get $U_{q1}$ and $U_{q2}$ upon which we do reduced-rank regression in the next step. 
    \item Compute the compact SVD of $U_{q1} = USV^T$ and define the operator $\tilde{A} = U^{T}U_{q2}VS^{-1}$.
    \item If desirable, dynamic modes are defined as $w = \lambda^{-1}U_{q2}VS^{-1}\tilde{w}$ for eigenvalues $\lambda$, eigenvectors $\tilde{w}$ of $\tilde{A}$.
\end{enumerate}
\vspace{-5mm}
\subsection{N4SID}
Numerical Algorithms for Subspace State Space System Identification (N4SID) \citep{van1994n4sid} utilizes a similar approach as the above to jointly estimate $A,B,C,D$ operators in a state space model from data $Y$ and $U$. Here we briefly describe the algorithm that we apply to estimate $A$ and $B$ that are used for comparison of partially observed systems, as first defined by \cite{n4sid}. We used code from \href{https://github.com/spmvg/nfoursid/tree/master}{https://github.com/spmvg/nfoursid/tree/master} for our implementation of n4sid. For the Subspace DMDc, we lift to a nonlinear space \textit{before} state estimation. 

For state estimation to succeed, standard conditions on the data state and input apply. In particular (1) the state vector is sufficiently excited (it explores all relevant dimensions of the state space), or the system is reachable, (2) the input sequence is persistently exciting, i.e., the Hankel matrix of the inputs is full rank, and (3) there is no linear state feedback, i.e. the state and the input are not collinear. Note that nonlinear feedback is permissible provided they are not collinear. Prediction in the SubSpaceDMDc is done with Kalman filtering, because state estimation must first take place.

Briefly, we explain the key computations behind N4SID. There are two slightly different approaches. The first algorithm is similar in spirit to Subspace DMD which we detail here: 
\subsubsection{Projection-Based N4SID}
As above, we create a Hankel data matrix of the observations, but also the input too, splitting this into past and future. First, we project out the data explained by $U_f$ in the future observations $Y_f$, but also the past observations and inputs $Z_p = \begin{bmatrix} U_p & Y_p \end{bmatrix}$, thereby removing its influence. Then to remove measurement and process noise biases, we project the future states onto the space explainable by the states and inputs in the past, $Z_p$. This yields our matrix $O = \Gamma X$,  which we split using SVD as before to get $\Gamma$, the extended observability matrix, and the states $X$ up to similarity. Noting again that our extended observability matrix has time-shifted structure, we can perform regression on the shifted components of $X$ given the instantaneous $U$, to arrive at $A,B$. The observability matrix $\Gamma$ also encodes $C$ in its top rows, which we can directly read out. However, we found this algorithm in practice to be less stable than the next one.

In pseudocode form, we have the following:

\begin{algorithm}[H]
\caption{Subspace DMD with Control (N4SID on lifted states)}
\label{alg:SubspaceDMDc}
\begin{algorithmic}[1]
\Require Output data $\mathbf{Y} \in \mathbb{R}^{p_{out} \times N}$, Input data $\mathbf{U} \in \mathbb{R}^{m \times N}$, past window $p$, future window $f$, system order $n$, regularization $\lambda$
\Ensure Estimated system matrices $\hat{\mathbf{A}}, \hat{\mathbf{B}}, \hat{\mathbf{C}}$
\vspace{0.5em}

\Procedure{BuildHankelMatrices}{$\mathbf{Y}, \mathbf{U}, p, f$}
    \State $T \gets N - p - f + 1$
    \State Construct Hankel matrices $\mathbf{Y}_p, \mathbf{U}_p, \mathbf{Y}_f, \mathbf{U}_f$
    \State $\mathbf{Z}_p \gets \begin{bmatrix} \mathbf{U}_p \\ \mathbf{Y}_p \end{bmatrix}$
    \State \Return $(\mathbf{Y}_f, \mathbf{U}_f, \mathbf{Z}_p, T)$
\EndProcedure
\vspace{0.5em}

\Procedure{ObliqueProjection}{$\mathbf{Y}_f, \mathbf{U}_f, \mathbf{Z}_p, \lambda, T$}
    \State $\mathbf{\Pi}_{\mathbf{U}_f^T}^{\perp} \gets \mathbf{I}_T - \mathbf{U}_f^T (\mathbf{U}_f \mathbf{U}_f^T + \lambda \mathbf{I})^{-1} \mathbf{U}_f$
    \State $\mathbf{Y}_{f, \perp} \gets \mathbf{Y}_f \mathbf{\Pi}_{\mathbf{U}_f^T}^{\perp}$
    \State $\mathbf{Z}_{p, \perp} \gets \mathbf{Z}_p \mathbf{\Pi}_{\mathbf{U}_f^T}^{\perp}$
    \State $\mathbf{O} \gets \mathbf{Y}_{f, \perp} \mathbf{Z}_{p, \perp}^{\dagger}$ \Comment{Oblique projection via pseudoinverse}
    \State \Return $\mathbf{O}$
\EndProcedure
\vspace{0.5em}

\Procedure{EstimateStateFromProjection}{$\mathbf{O}, n$}
    \State $\mathbf{U}_o, \mathbf{S}_o, \mathbf{V}_o \gets \mathrm{SVD}(\mathbf{O})$
    \State Truncate to rank $n$: $\mathbf{U}_n, \mathbf{S}_n, \mathbf{V}_n$
    \State $\hat{\mathbf{\Gamma}}_f \gets \mathbf{U}_n \sqrt{\mathbf{S}_n}$ \Comment{Estimated observability matrix}
    \State $\hat{\mathbf{X}} \gets \sqrt{\mathbf{S}_n} \mathbf{V}_n^T$ \Comment{Estimated state sequence}
    \State \Return $(\hat{\mathbf{\Gamma}}_f, \hat{\mathbf{X}})$
\EndProcedure
\vspace{0.5em}

\State $\mathbf{Y}_f, \mathbf{U}_f, \mathbf{Z}_p, T \gets \textsc{BuildHankelMatrices}(\mathbf{Y}, \mathbf{U}, p, f)$
\State $\mathbf{O} \gets \textsc{ObliqueProjection}(\mathbf{Y}_f, \mathbf{U}_f, \mathbf{Z}_p, \lambda, T)$
\State $\hat{\mathbf{\Gamma}}_f, \hat{\mathbf{X}} \gets \textsc{EstimateStateFromProjection}(\mathbf{O}, n)$
\State
\State \Comment{Align data for regression}
\State $\hat{\mathbf{X}}_{\text{current}} \gets \hat{\mathbf{X}}[:, 0:T-1]$
\State $\hat{\mathbf{X}}_{\text{next}} \gets \hat{\mathbf{X}}[:, 1:T]$
\State $\mathbf{U}_{\text{mid}} \gets \mathbf{U}[:, p : p+T-1]$
\State
\State \Comment{Solve for system matrices}
\State $\begin{bmatrix} \hat{\mathbf{A}} & \hat{\mathbf{B}} \end{bmatrix} \gets \hat{\mathbf{X}}_{\text{next}} \begin{bmatrix} \hat{\mathbf{X}}_{\text{current}} \\ \mathbf{U}_{\text{mid}} \end{bmatrix}^{\dagger}$
\State $\hat{\mathbf{C}} \gets$ first $p_{out}$ rows of $\hat{\mathbf{\Gamma}}_f$
\State
\State \Return $\hat{\mathbf{A}}, \hat{\mathbf{B}}, \hat{\mathbf{C}}$
\end{algorithmic}
\end{algorithm}

\section{Misaligned Input Spaces}\label{app: misalignedinput}


For any orthogonal matrix \(C\), the following equivalence holds:
\begin{align}\label{eq:C_eq}
  y = Cx 
  \Longleftrightarrow 
  \dot{y} = CA_{1}C^{T}\,x + C B_{1}\,u(t)
\end{align} 
Now consider the case where inputs are not equivalent in each system, but that they are also related by a coordinate transform: $$u_y(t) = C_u u_x(t)$$ Then Eq. \ref{eq: controllability_eqn} resolves to:
\begin{equation}
      \dot{y} = CA_{1}C^{T}\,x + C B_{1}C_u\,u_y(t)
\end{equation}
This motivates the dissimilarity metric that seeks to jointly optimize $C$ and $C_u$, with the second term in Eq. \ref{eq: \method} generalizing to 
$$\min_{C_u \in O(n)}||CB_1C_u - B_2|| $$

when $\alpha = 0$, \eqref{eq: \method} is the so-called two-sided Procrustes problem, which when solved jointly resolves to comparing the singular values of $B_1,B_2$: $||\Sigma_1 - \Sigma_2||$, which can be computed efficiently. When $\alpha \neq 0$, the two minimizations need to be jointly optimized. The method of optimization from \cite{ostrow2024beyond} can be effectively generalized to do so, with note that this is a larger optimization problem and requires longer optimization time (but see next section). 


If the inputs that are directly applied to the system are known, as in RNN or RL models ($\dot{x} = f(x,u)$), this joint optimization procedure can be discarded. Likewise, when the inputs are aligned in time, Procrustes or other spatial alignment methods can be directly applied to the inputs first. Note that this input comparison does not directly compare the dynamics of the input, but rather how the input is read into the system. If one is interested in comparing the dynamics of the input as well, then DSA can be run on the input directly. 

\section{Solving for optimal orthogonal $C$ efficiently} \label{app: fastCcalc}

The \method formulation allows for efficient solving of the optimal $C \in O(n)$. Recall that 

\begin{align}
    DSA(A_x,A_y) = \min_{C\in O(n)} \left||A_x - CA_yC^T\right||_F^2
\end{align}

Is a non-convex optimization problem, and hence has to be solved iteratively \cite{ostrow2024beyond}. However, the addition of the control constraint, $\left||B_x - CB_y\right||_F^2$ means that we can solve this problem using convex optimization for $\alpha = 0.5$. Observe that under similarity, 

$$\tilde{A} = CAC^T, \tilde{B} = CB \implies \tilde{A}\tilde{B} = CAB$$

This suggests that we can identify $C$ via Procrustes alignment on the controllability matrix $K = \begin{pmatrix} B & AB & A^2 B & \dots A^nB \end{pmatrix}$, where $A \in \mathbb{R}^{n \times n}$:

\begin{align} \label{eq: controllability_eqn}
\min_{C \in O(n)} \left||K_1 - CK_2\right||_F^2
\end{align}

The minimizer $C^*$ has a closed-form solution via orthogonal Procrustes. Likewise, jointly aligning the input dimension via $C_u$ (Appendix Sec. \ref{app: misalignedinput}) can be done in closed form as well via the two-sided Procrustes solution. This results in an acceleration of multiple orders, with the computation of $C$ taking O(1 millisecond), as opposed to O(1 second). 

However, this formulation can result in $C^*$ that are biased towards the more controllable directions, i.e. $B$ can have an inordinate effect or can dominate. In practice, we found that using this approach with a ground truth $C$ resulted in the state similarity score becoming biased near dimension $30$ (that is, $A \in \mathbb{R}^{30 \times 30}$). While this is still quite large, and the biases are small (average deviation $O(0.01)$ per element), we can do better. We can add further constraints to $C$, by noticing that $A^T$ also holds in the previous implication under similarity:

$$\tilde{A} = CAC^T, \tilde{B} = CB \implies \tilde{A}^T\tilde{B} = CA^TB$$
Thus, we can concatenate these powers as well to K, giving:
$$K = \begin{pmatrix} B & AB  & A^TB & \dots & A^nB &  A^{T^n} B \end{pmatrix}$$

Where the metric is once again Eq. \ref{eq: controllability_eqn}. This improves the optimization stability on $A$ until at least dimension $150$ for O(0.001) error per element error, which is more than enough in practice for \method. We have the following lemma which states that this metric captures equivalency between two linear systems. 

\begin{lemma}
    Given two linear systems $x_{t+1} = A_x x_t + B_x u_t $ and  $y_{t+1} = A_y y_t + B_y u_t$, Eq. \ref{eq: controllability_eqn} is equal to zero \textbf{if and only if} $y = Cx$ for some $C^TC = I$.
\end{lemma}
\begin{proof}
    Let us first consider the forward direction. Assuming $y=C^*x$, then we have the equivalence relationships $A_x = C^{*^T}A_yC^*$ and $B_x = C^{*^T}B_y$. Applying this relationship to $K_x$, we have
    \begin{align*}
        K_x &= \begin{pmatrix} C^{*^T} B_y & C^{*^T}A_yB_y & C^{*^T}A_y^TB_y & \dots & C^{*^T}A_y^n B_y \end{pmatrix} \\
        &= C^{*^T}K_y
    \end{align*}
    For which $\min_{C \in O(n)} \left||K_x - CK_y\right||_F^2 = 0$ evidently at $C^*$. 

    Now consider the reverse direction. We can expand Eq. \ref{eq: controllability_eqn}
 as: 
\begin{align*}
     \left||K_x - CK_y\right||_F^2 &= \left||B_x - CB_y\right||_F^2 + \left||A_xB_x - CA_yB_y\right||_F^2 + \left||A_x^TB_x - CA_y^TB_y \right||_F^2 + \dots
\end{align*}
 
    For $\min_{C \in O(n)} \left||K_x - CK_y\right||_F^2 = 0$, each subterm must be zero for minimizer $\tilde{C}$. This immediately gives $B_x = \tilde{C}B_y$. Inspecting the next term, we substitute this relationship, giving
    \begin{align}
       0 =  ||A_xB_x - \tilde{C}A_yB_y||_F^2  &= ||A_x\tilde{C}B_y - \tilde{C}A_yB_y||_F^2 \\
        &= ||(A_x\tilde{C}- \tilde{C}A_y)B_y||_F^2 \\
        &\implies A_x\tilde{C} = \tilde{C}A_y \\
    \end{align}
With the last step following from $B_y \neq 0$ . This in turn gives $A_x = \tilde{C}A_y\tilde{C}^T$. We can similarly apply this reasoning to the next expression, which gives the same result. Reversing the previous logic, we have $x_{t+1} = \tilde{C}A_y\tilde{C}^Tx_t + \tilde{C}B_yu_t \implies y= \tilde{C}^Tx$.
\end{proof}

For a given $A,B$, the standard right Procrustes problem is written as:

\begin{align}
C^* &= \text{argmin}_{C \in O(n)} \left||CA - B\right||_F^2 \\
&= \text{argmax}_{C \in O(n)} <CA,B>_F = \text{Tr}[(CA)^TB]
\end{align}
Writing the form of this problem with $K_1,K_2$, we can separate out individual elements in the Frobenius inner product, giving

\begin{align}
    <CK_1,K_2>_F &= \sum_{i=0}^n <CA_x^i B_x, A_y^iB_y>_F + <CA_x^{T^i}B_x, A_y^{T^i}B_y>_F \\
    &= \sum_{i=0}^n <C, A_y^i B_y(A_x^i B_x)^T>_F + <C, A_y^{T^i}B_y(A_x^{T^i}B_x)^T>_F \\
    &= <C, \sum_{i=0}^n A_y^iB_y(A_x^iB_x)^T + A_y^{T^i}B_y(A_x^{T^i}B_x)^T>_F
    \end{align}

With the last steps due to linearity of the inner product and the second step using the trace permutation identity. This gives the maximum over $C \in O(n)$ to be

$$C^* = UV^T \qquad \text{where}   \quad \sum_{i=0}^n A_y^iB_y(A_x^iB_x)^T + A_y^{T^i}B_y(A_x^{T^i}B_x)^T= USV^T$$

In practice, taking large matrices $A$ to many powers results in numerical instability issues, especially when $\lambda_{max}(A) > 1$. Algorithmically, we check the condition number of $A^n$ before choosing to include the term in the controllability matrix. If it is too small or too large, we stop.

\subsection{Generalizing the Wasserstein Distance for \method} \label{app: Wasserstein}

Recall the Wasserstein distance over DMD eigenvalues, 

\begin{equation}\label{eq: redDSA2}
\text{DSA}(\Lambda_1,\Lambda_2)  := \, \min_{P \in \Pi(n) } ||P\Lambda_1P^T - \Lambda_2||_F 
\end{equation}
This metric respects the notion of equivalency under general similarity transforms, $A \rightarrow CAC^{-1}$ for invertible C's, given that only eigenvalues are preserved under these transformations. We would like to identify a similar metric for input driven systems. To motivate our metric, consider applying a diagonalizing transform to the dynamics of our input-driven system:

\begin{align}
    x_{t+1} &= Ax_t + Bu_t \\
    &= V\Lambda V^{-1} x_t + Bu_t \\
    V^{-1}x_{t+1} &= \Lambda V^{-1}x_t + V^{-1}B u_t
\end{align}
We observe that the corresponding feature of the input to each eigenvalue $\lambda$ is the row vectors on $V^{-1}B$, which we henceforth term the eigenmode-input interaction matrix. We can easily show that these features are invariant to any invertible transform. Given a transform $A \rightarrow CAC^{-1}, B \rightarrow CB$, 
\begin{align}
    V\Lambda V^{-1} \rightarrow CV \Lambda V^{-1} C^{-1} := \tilde{V}\Lambda \tilde{V}^{-1}
\end{align}
Hence $V \rightarrow CV$
\begin{align}
    V^{-1}B \rightarrow (CV)^{-1}CB = V^{-1}C^{-1}CB = V^{-1}B
\end{align}

Thus, a natural extension to Eq. \ref{eq: redDSA2} is the joint Wasserstein distance over $[\Lambda_i, (V^{-1}B)_i]$. Denoting $\Lambda^1$ the set of eigenvalues for system one, and denoting $\pi$ a permutation map, 

\begin{align}
    \text{InputDSA}(\Lambda^1, \Lambda^2, V_1^{-1}B_1, V_2^{-1}B_2, \alpha) = \min_{\pi} \sum_i [\alpha(\Lambda_i^1 - \Lambda_{\pi(i)}^2)^2 +  (1-\alpha)|(V_1^{-1}B_1)_{i} -(V_1^{-1}B_1)_{\pi(i)}|_2^2 ]
\end{align}
This metric is intuitive: $V^{-1}B$ describes how input direction interacts with the independent eigenmodes, which is related to the controllability of that mode. However, this metric has numerical stability issues. First, eigenvalues can only be identified up to an arbitrary phase-hence, we are forced to study instead the norms of each eigenmode-input interaction, $|(V_1^{-1}B_1)_i|_2$. This loses information but works reasonably for small systems. Identifying the eigenvectors of an arbitrary matrix is challenging for poorly-conditioned matrices. Hence, we suggest evaluating the conditioning of the DMD matrix before applying this metric. 

\section{\method pseudocode}
\begin{algorithm}
\caption{\method}
\label{alg:\method}
\begin{algorithmic}[1]
\Require $X_1, X_2 \in \mathbb{R}^{n \times t \times d}, \; U_1, U_2 \in \mathbb{R}^{n \times t \times \ell}, \;\text{number of delays } q, \text{nonlinear lifting functions } \phi_1, \phi_2$ \\
\hspace{1.2em} $\; \text{rank for state-space } r,$ 
\Ensure Similarity transform distance $d$ between the two dynamical systems
\vspace{0.5em}

\State $A_1, B_1 \gets \textsc{SubspaceDMDc}(\phi_1(X_{_1}), \phi_2(U_{1}), r)$
\State $A_2, B_2 \gets \textsc{SubspaceDMDc}(\phi_1(X_{2}), \phi_2(U_{2}), r)$

\State $d = \min_{\substack{C \in O(n) \\ C_u \in O(n) \\ }}
  \alpha \, \| C A_1 C^\top - A_2 \|_F
  + (1-\alpha) \, \| C B_1 C_u - B_2 \|_F$

\State \Return $d$
\end{algorithmic}
\end{algorithm}

\begin{algorithm}
\caption{Dynamic Mode Decomposition with Control (DMDc)}
\label{alg:DMDc}
\begin{algorithmic}[1]
\Require Delay-embedded states $H_X$, inputs $H_U$, truncation ranks $r_{\text{all}}, r_{\text{state}}$, ridge regularization $\lambda$
\Ensure Dynamics operator $A$ and $B$ 
\vspace{0.5em}

\Procedure{BuildSnapshots}{$H_X, H_U$}
  \State $X_- \gets H_X[:, 1\!:\!-1]$, \quad $X_+ \gets H_X[:, 2\!:\!]$
  \State $U_- \gets H_U[:, 1\!:\!-1]$
  \State $\Omega \gets \begin{bmatrix} X_- \\ U_- \end{bmatrix}$ 
  \State \Return $(X_+, X_-, U_-, \Omega)$
\EndProcedure
\vspace{0.5em}

\Procedure{SVDs}{$X_+, \Omega$}
  \State $(U_p, \Sigma_p, V_p) \gets \mathrm{SVD}(\Omega)$
  \State Partition $U_p = \begin{bmatrix} U_{p1} \\ U_{p2} \end{bmatrix}$ into state/input blocks
  \State $(U_r, \Sigma_r, V_r) \gets \mathrm{SVD}(X_+)$
  \State \Return $(U_{p1}, U_{p2}, \Sigma_p, V_p, U_r)$
\EndProcedure
\vspace{0.5em}

\Procedure{ReduceRank}{$U_{p1}, U_{p2}, \Sigma_p, V_p, U_r$}
    \State Truncate to\textbf{ $r_{\text{all}}$:} $U_{p1}\!\leftarrow\!U_{p1}[:,1\!:\!r_{\text{all}}]$, $U_{p2}\!\leftarrow\!U_{p2}[:,1\!:\!r_{\text{all}}]$, $V_p\!\leftarrow\!V_p[:,1\!:\!r_{\text{all}}]$, $\Sigma_p\!\leftarrow\!\Sigma_p[1\!:\!r_{\text{all}}]$
      \State Truncate to\textbf{ $r_{\text{state}}$:} $U_r \!\leftarrow\! U_r[:,1\!:\!r_{\text{state}}]$
     \State \Return $(U_{p1}, U_{p2}, \Sigma_p, V_p, U_r)$
    \EndProcedure
    \vspace{0.5em}
    
\Procedure{ComputeOperators}{$X_+, V_p, \Sigma_p, U_{p1}, U_{p2}, U_r, \lambda$}
  \State $\Sigma_p^\dagger(\lambda) \gets \mathrm{diag}\!\left(\frac{\sigma_i}{\sigma_i^2+\lambda}\right)$
  \State $A \gets X_+ \, V_p \, \Sigma_p^\dagger(\lambda) \, U_{p1}^\top$
  \State $B \gets X_+ \, V_p \, \Sigma_p^\dagger(\lambda) \, U_{p2}^\top$
\State Project to the state space $\tilde A \gets U_r^\top A U_r$, \quad $\tilde B \gets U_r^\top B$
\Return $(\tilde A, \tilde B)$
\EndProcedure
\vspace{0.5em}

\State $X_+, X_-, U_-, \Omega \gets \textsc{BuildSnapshots}(H_X, H_U)$
\State $U_{p1}, U_{p2}, \Sigma_p, V_p, U_r \gets \textsc{SVDs}(X_+, \Omega)$
\State $U_{p1}, U_{p2}, \Sigma_p, V_p, U_r \gets \textsc{ReduceRank}(\cdot)$
\State $A, B \gets \textsc{ComputeOperators}($$X_+, V_p, \Sigma_p, U_{p1}, U_{p2}, U_r, \lambda$)
\State \Return $A, B$
\end{algorithmic}
\end{algorithm}

\section{Hyperparameter tuning for \method}\label{app: hpsweep}

\paragraph{Delay} 
In \method, the delay parameter controls the size of the delay embedding used to estimate the dynamics operator. 
If too few delays are chosen, the embedding may distort the data and amplify noise. 
Conversely, too many delays fold the data into unnecessarily high dimensions, making it more difficult to model the dynamics with DMD \citep{ostrow2024delay}. 

\paragraph{Rank} 
SubspaceDMDc involves one rank parameter $r$, corresponding to the dimensionality of the latent state space. In practice, selecting an $r$ slightly higher than the true state dimension often yields a better estimation of the $A$ matrix. 

\textbf{Hyperparameter tuning pipeline} We suggest a two-step hyperparameter tuning pipeline for \method: jointly optimize the delay and the total rank $p$ jointly as the first step, followed by optional refinement of $r$ to obtain a more accurate estimate of the system's dynamics. We recommend jointly selecting the delay length according to the following criteria:

\begin{itemize}
    \item \textbf{Prediction accuracy:} 
    The delay embedding should enable accurate modeling of the dynamics. 
    To evaluate this, we split the dataset into training and test sets, fit \method (via SubspaceDMDc) on the training set, and assess performance on the test set using the mean absolute standardized error (MASE), a standard metric for time-series forecasting. 
    MASE compares the forecast error of the model against that of a naïve persistence baseline predictor and is defined as
    \[
    \text{MASE} = 
    \frac{\tfrac{1}{T} \sum_{t=1}^{T} \lvert y_t - \hat{y}_t \rvert}
         {\tfrac{1}{T-1} \sum_{t=2}^{T} \lvert y_t - y_{t-1} \rvert}.
    \]
    A value $\text{MASE} < 1$ indicates that DMDc predicts next-step activity (using the estimated operators $A$ and $B$) more accurately than simply copying the current time step.
    
    \item \textbf{Model complexity:} 
    The estimated operators $A$ and $B$ should not be overly complex or dominated by spurious features (e.g., many small eigenvalues clustered near zero). 
    To assess this, we compute the Akaike Information Criterion (AIC) for next-step prediction on the test set. 
    AIC balances predictive accuracy against model complexity and, in our setting, is given by
    \[
    \text{AIC} =
    \ln \!\left( \frac{1}{N} \sum_{j=1}^{N} (x_j - y_j)^2 \right)
    + \frac{2(r^2 + 1)}{N}.
    \]
\end{itemize}
Overall, we aim to select a rank that is small enough to avoid inflating the AIC, while still yielding good predictive accuracy (i.e., low MASE). After joint optimization of the delay and rank $p$, we can optionally fine-tune $r$ by gradually decreasing its value until the MASE increases sharply. 

\section{Partially Observed System Comparison Further Detail} \label{app: po_comp}

We discretely simulated the following equations (repeated from \ref{eq: nonlin_po}): 

\begin{align}
    x_{t+1} &= A(x_t + gF\tanh(x_t)) + B(u_t + \tanh(u_t)) \\
    y_t &= \begin{pmatrix} \mathbf{I}_{d} & \mathbf{0}_{n - d} \end{pmatrix}x_t + \epsilon_t
\end{align}

We generated two matrices $A_1, A_2$, sampling each element i.i.d. from a standard normal distribution. To enforce stability of these matrices, we globally rescaled the matrices by a term $\rho/\lambda_{max}$, where $\lambda_{max}$ is the max eigenvalue of the sampled matrix and $0 < \rho < 1$. We arbitrarily picked $\rho_1 = 0.92$ and $\rho_2 = 0.82$ to ensure a significant difference in the intrinsic dynamics, but not so large as to make the data obviously different. 
We set $g = 0.1$ for each system, and fixed $F$ to be the matrix defined as $F_{ij} = \delta_{ij}\delta_{i \leq d}$ where $d$ is the number of observed states in the observation matrix $C = \begin{pmatrix} \mathbf{I}_{d} & \mathbf{0}_{n - d} \end{pmatrix}$. 
We sampled $B_1,B_2$ from normal distributions as well, with $B_{1_{ij}} \sim N(0, g_1)$, $B_{2_{ij}} \sim N(0,g_2)$, setting $g_1 = 0.5,g_2 =2.0$. 
We sampled $\epsilon_i  \sim N(0,0.01)$ for each observed index for each time-point.

Across Figs. \ref{fig:fig2}b,c,d,e, we simulated 20-dimensional systems with only 2 dimensions observed, for 5000 timepoints. For every type of DMD, we applied delay embeddings of size 150, and fit state space / dynamics matrices with rank 20. We chose these parameters by inspecting the spectral distribution of the estimated observability matrix (line 16 of Algorithm \ref{alg:SubspaceDMDc}) across multiple delays. We added delays under the largest modes before the spectral drop-off point stopped changing (similar to the idea of a false neighbors analysis, \citealt{kennelfalseneighbors}), then picked the elbow of that curve. We selected the maximum of those values for each of the four systems. We observe these curves in Fig. \ref{app: fig: spectral_curves}. However, we note that \method is robust to a number of different ranks (Fig. \ref{app: fig: ranksweep}), both larger and smaller than the true system size. 

We computed silhouette score using Scikit-Learn on the precomputed \method distance. Based on some given label (here, state or input ground-truth similarity), the dataset is divided into subsets $C_1, C_2, \ldots C_n$  with each data point $x_1, x_2, ...x_N$ belonging to one subset. Define the labels (cluster index) of each point as $c_1, c_2, ... c_N$. Next, the mean intra-cluster and the minimum mean inter-cluster distance is computed for each data point:

\begin{align*}
    a(i) &= \frac{1}{|C_{c(i)}|}\sum^N_{C_{c(j)} = C_{c(i)}, i \neq j} d(x_i,x_j) \\
    b(i) &= \min_{j \neq c(i)} \frac{1}{|C_j|}\sum^N_{c(k) = j} d(x_i,x_k)
\end{align*}

Where $|C_{c(i)}|$ denotes the cardinality of the set, and $d(\cdot, \cdot)$ denotes the distance function to be used. In our setting, we use the \method input or state distances for $d$. Lastly, the silhouette score is computing as:

\begin{align*}
    S  = \frac{1}{N}\sum_{i=1}^N \frac{b(i) - a(i)}{\max(a(i),b(i))}
\end{align*}

The silhouette score approaches 1 when all points in each class are strongly separated and there is minimal distance between the points within each class, while it is 0 if the inter- and intra-cluster distances are equivalent. It is notable that a silhouette score of $~0.7$ can correspond to perfect linear classification of all classes, as deviations from $1.0$ can be caused by within-class variance that remains non-overlapping with other classes. 

\begin{figure}
    \centering
    \includegraphics[width=0.65\linewidth]{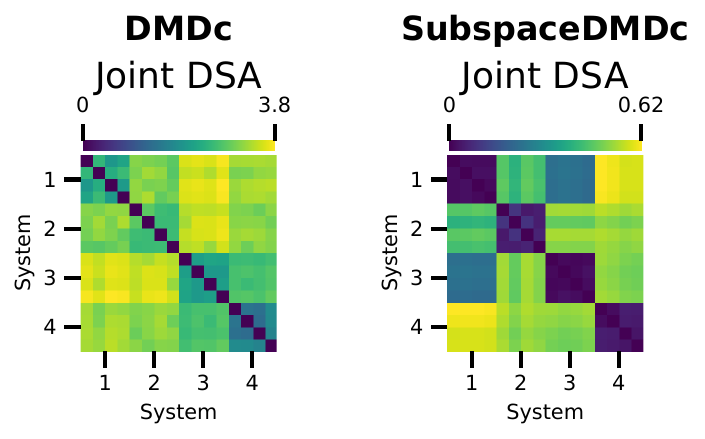}
    \caption{\textbf{Joint \method Comparison using DMDc and SubspaceDMDc} The sum of jointly-optimized state and input distances is presented here, with $\alpha=0.5$. Comparisons were generated on the same dataset as in Fig. \ref{fig:fig2}c and d. DMDc Silhouette score on state, input: 0.235, 0.088. SubspaceDMDc Silhouette score on state, input: 0.368, 0.55.}
    \label{app: fig: jointdsa}
\end{figure}

\begin{figure}
    \centering
    \includegraphics[width=0.85\linewidth]{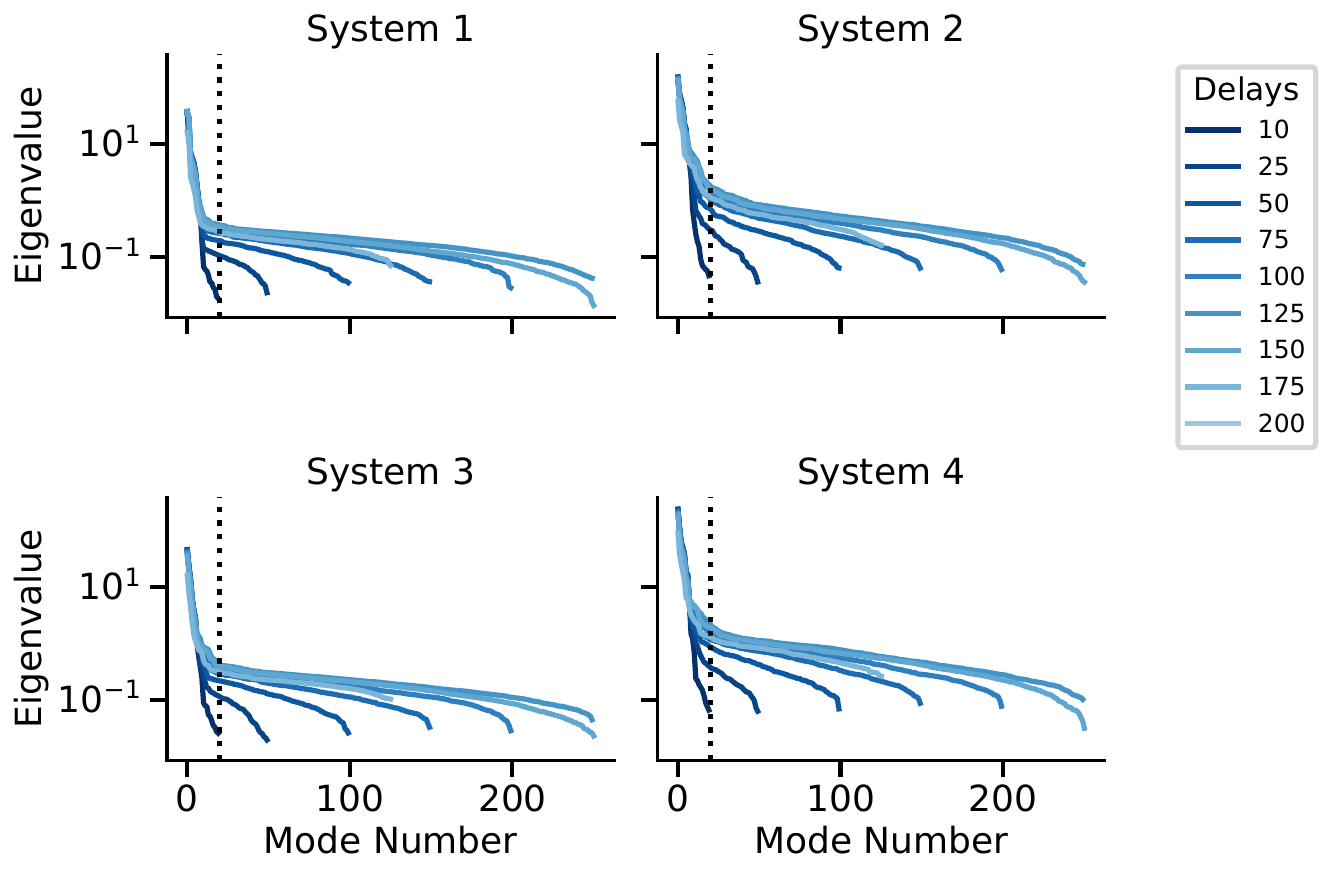}
    \caption{Spectral content of the Extended Observability Matrix from Subspace DMDc for each system in Fig. \ref{fig:fig2}b across multiple delays. Dotted line indicates rank 20.}
    \label{app: fig: spectral_curves}
\end{figure}
\begin{figure}
    \centering
    \includegraphics[width=0.95\linewidth]{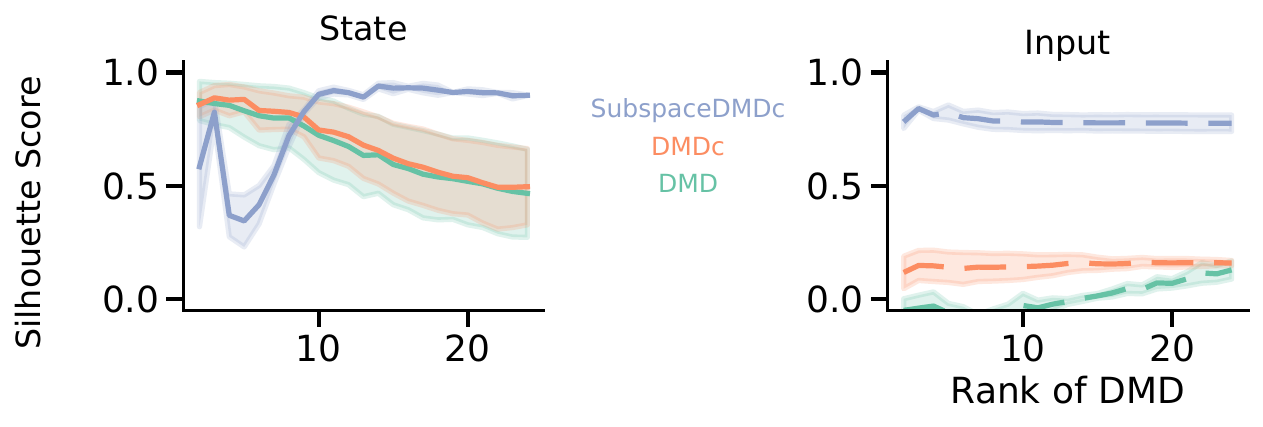}
    \caption{Effect of rank in each DMD algorithm on clustering scores in each method, utilizing 20 dimensional systems with 2 dimensions observed, and 1000 datapoints per dataset.}
    \label{app: fig: ranksweep}
\end{figure}

\section{Input Noise Generation}
\label{app:input_noise}

\begin{figure*}[!tbhp]
    \centering
    \includegraphics[width=\linewidth]{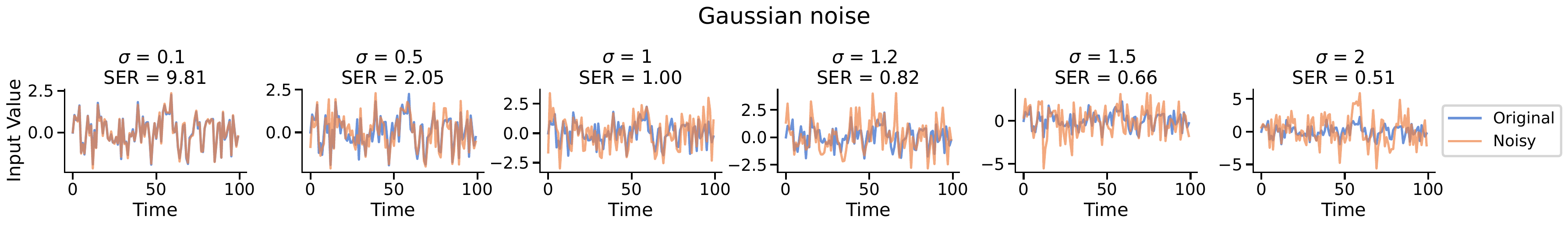}
    \includegraphics[width=\linewidth]{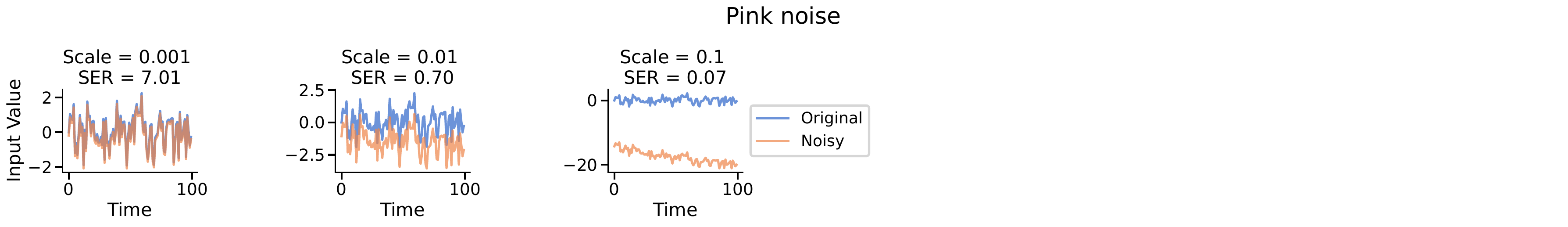}
    \includegraphics[width=\linewidth]{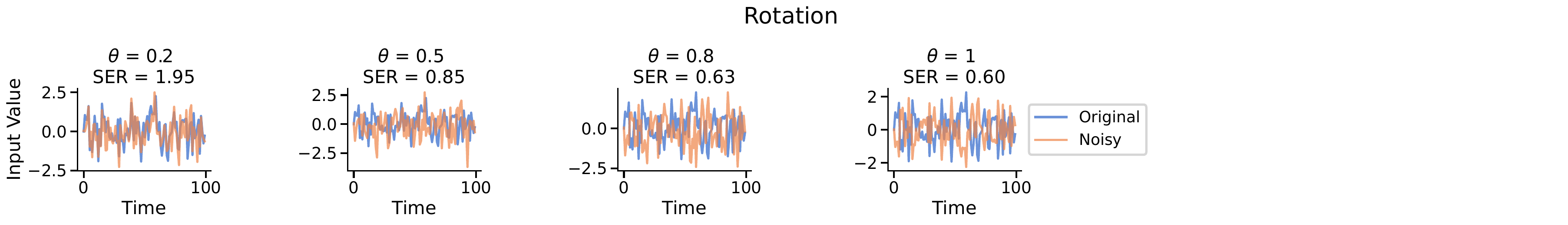}
    \includegraphics[width=\linewidth]{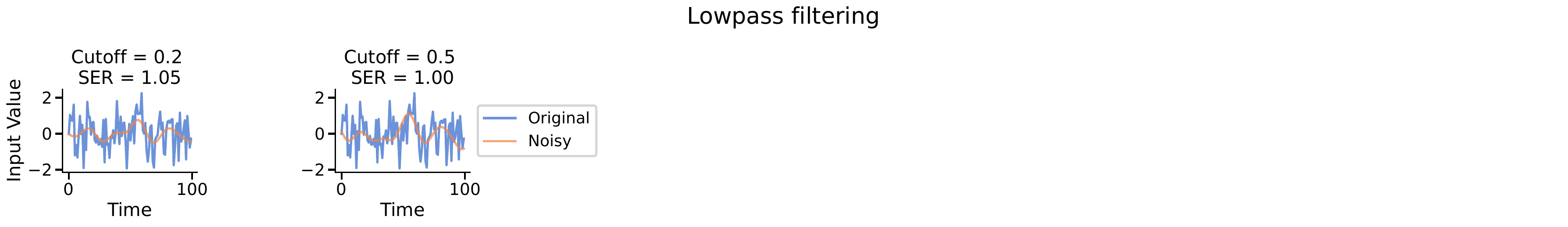}
    \includegraphics[width=\linewidth]{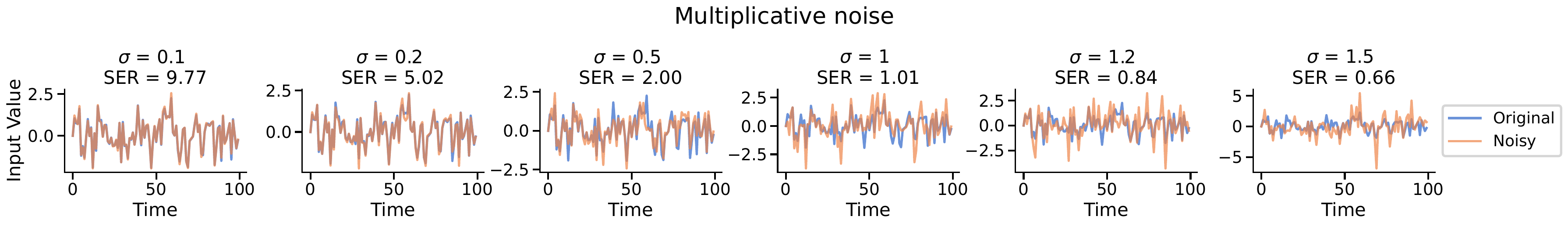}
    \caption{Effect of different noise types and levels on the input time series of the nonlinear dynamical system as in Fig. \ref{fig:fig2}. The plots show activity along four example observation dimensions received by the networks during an example trial. Here, we show gaussian white noise, pink noise, rotation, low-pass filter, and multiplicate noise applied to the input.}
    \label{app:noise_visualization_part1}
\end{figure*}

\begin{figure*}[!tbhp]
    \centering
    \includegraphics[width=\linewidth]{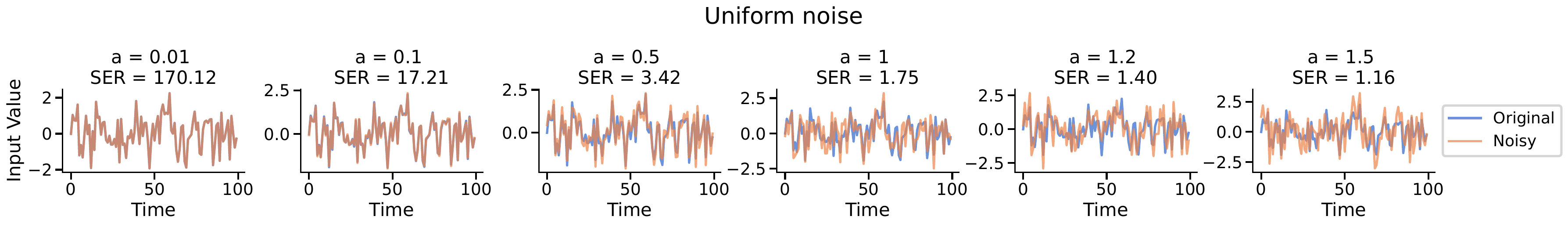}
    \includegraphics[width=\linewidth]{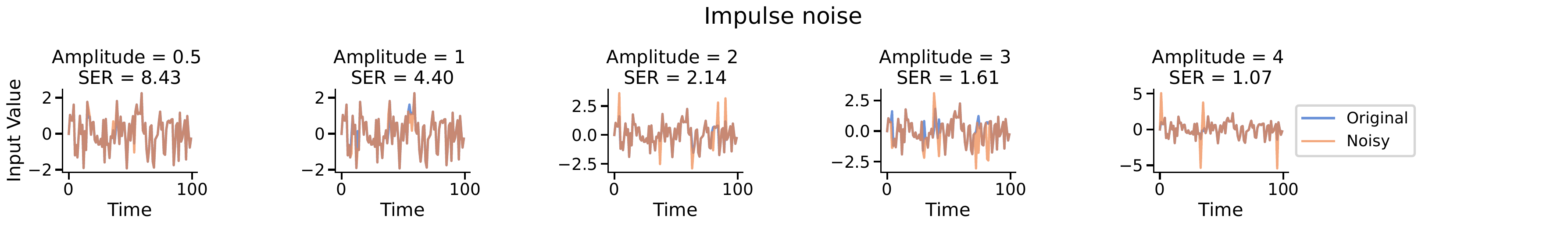}
    \includegraphics[width=\linewidth]{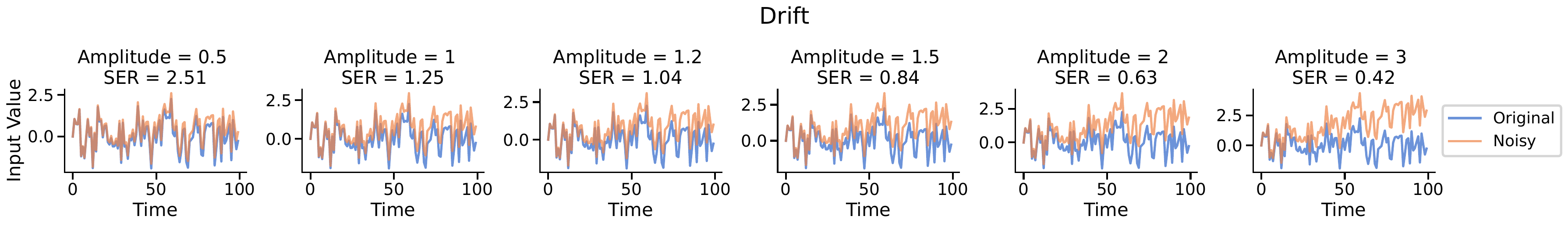}
    \includegraphics[width=\linewidth]{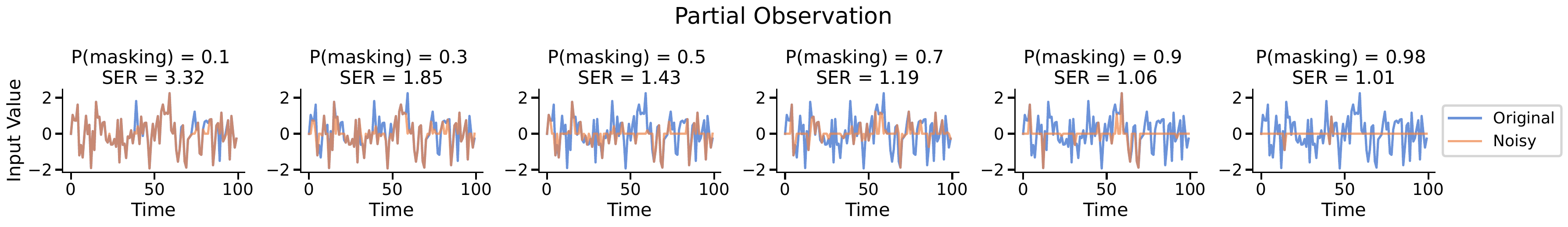}
    \includegraphics[width=\linewidth]{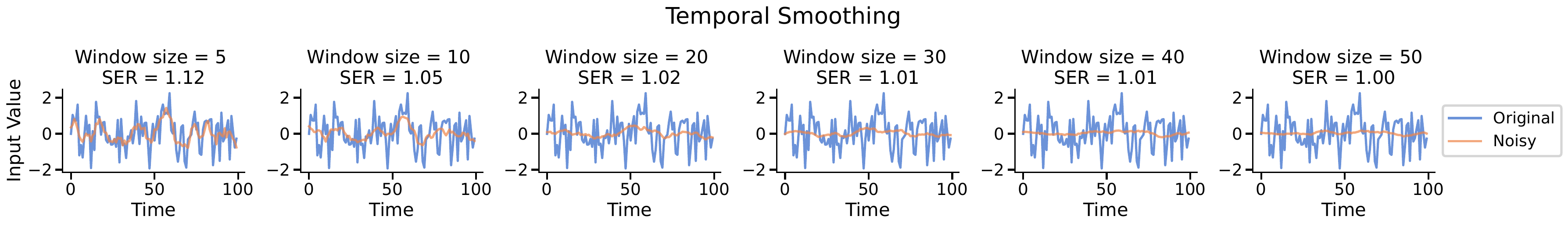}
    \caption{Effect of different noise types and levels on the input time series of the nonlinear dynamical system as in Fig. \ref{fig:fig2}. The plots show activity along four example observation dimensions received by the networks during an example trial. Here, we show noise sampled from an uniform range, impulse noise, random drift, partial observability, and temporal smoothing applied to the input.}
    \label{app:noise_visualization_part2}
\end{figure*}

To assess the robustness of \method to noisy or corrupted inputs, we systematically added different types of noise or transformations to the input time series of the nonlinear dynamical systems we created. Below we describe how each type of noise was generated. 
We visualize examples of the noise-corrupted input in Fig. \ref{app:noise_visualization_part1} and Fig. \ref{app:noise_visualization_part2}
. 
\paragraph{Gaussian (white) noise.} White Gaussian noise was added independently to each input channel:
\[
\tilde{u}(t) = u(t) + \mathcal{N}(0, \sigma^2),
\]
where $\sigma$ is set by the noise level.

\paragraph{Pink noise.} Pink ($1/f$) noise was generated in the frequency domain with power spectrum proportional to $1/f^{\alpha}$ (with $\alpha = 1$ by default), then inverse Fourier transformed and scaled to the desired amplitude.

\paragraph{Rotation.} For two-dimensional input signals, we applied a random planar rotation:
\[
\tilde{u}(t) = R(\theta)\, u(t), \quad 
R(\theta) = 
\begin{bmatrix}
\cos \theta & -\sin \theta \\
\sin \theta & \cos \theta
\end{bmatrix},
\]
with rotation angle $\theta$ proportional to the noise level.

\paragraph{Low-pass filtering.} Inputs were smoothed using a digital Butterworth low-pass filter with cutoff frequency set by the noise level. Larger values corresponded to stronger filtering.

\paragraph{Multiplicative noise.} Each input channel was scaled by a random Gaussian factor:
\[
\tilde{u}(t) = u(t) \cdot \eta, \quad 
\eta \sim \mathcal{N}(1, \sigma^2),
\]
where $\sigma$ is set by the noise level.

\paragraph{Uniform noise.} Additive noise sampled uniformly from $[-a, a]$ was added to each channel, where $a$ is the noise level.

\paragraph{Impulse noise.} At each time point, with probability $p$, an impulse of magnitude $\pm \alpha$ (set by the noise level) was added to the input.

\paragraph{Baseline drift.} A slow oscillatory drift was added to each channel:
\[
d(t) = A \sin(2 \pi f t) + \tfrac{A}{2} \sin(4 \pi f t),
\]
where $A$ is the drift amplitude (noise level) and $f$ is a low drift rate.

\paragraph{Partial observability.} A random fraction of input time series was masked with zeros, with masking probability given by the noise level.

\begin{table}[h]
\centering
\caption{Noise levels used in experiments for each noise type.}
\begin{tabular}{ll}
\toprule
Noise type & Levels used \\
\midrule
White (Gaussian) & 0.1, 0.5, 1, 1.2, 1.5, 2, 3 \\
Pink & 0.01, 0.1, 0.2, 0.3, 0.5, 0.7, 0.9, 1\\
Rotation & 0.2, 0.5, 0.8, 1 \\
Low-pass & 0.2, 0.5 \\
Multiplicative & 0.1, 0.2, 0.5, 1, 1.2, 1.5, 2, 3 \\
Uniform & 0.01, 0.1, 0.5, 1, 1.2, 1.5, 2, 3 \\
Impulse & 0.5, 1, 2, 3, 4 \\
Baseline drift & 0.5, 1, 1.2, 1.5, 2, 3, 4 \\
Partial observability & 0.1, 0.3, 0.5, 0.7, 0.9, 0.98 \\
Temporal smoothing & 5, 10, 20, 30, 40, 50 \\
\bottomrule
\end{tabular}
\end{table}

\newpage
\section{Ordinary Least Squares Biases Estimates of $A$ in the presence of input noise}
\label{app:noise_effect}
Using two datasets: the nonlinear dynamical systems as in Fig. \ref{fig:fig2}, and the RNNs trained on Random Target Task as in Fig. \ref{fig:fig3}, we show that increasing the noise variance systematically contracts the singular spectrum of $B$ toward zero. Because the true input effect is underfit, the SubspaceDMDc regression inflates the real part of $A$'s eigenvalues to absorb the variance in the inputs that correlates with the state.

\begin{figure*}[h] 
    \centering
    \includegraphics[width=\linewidth]{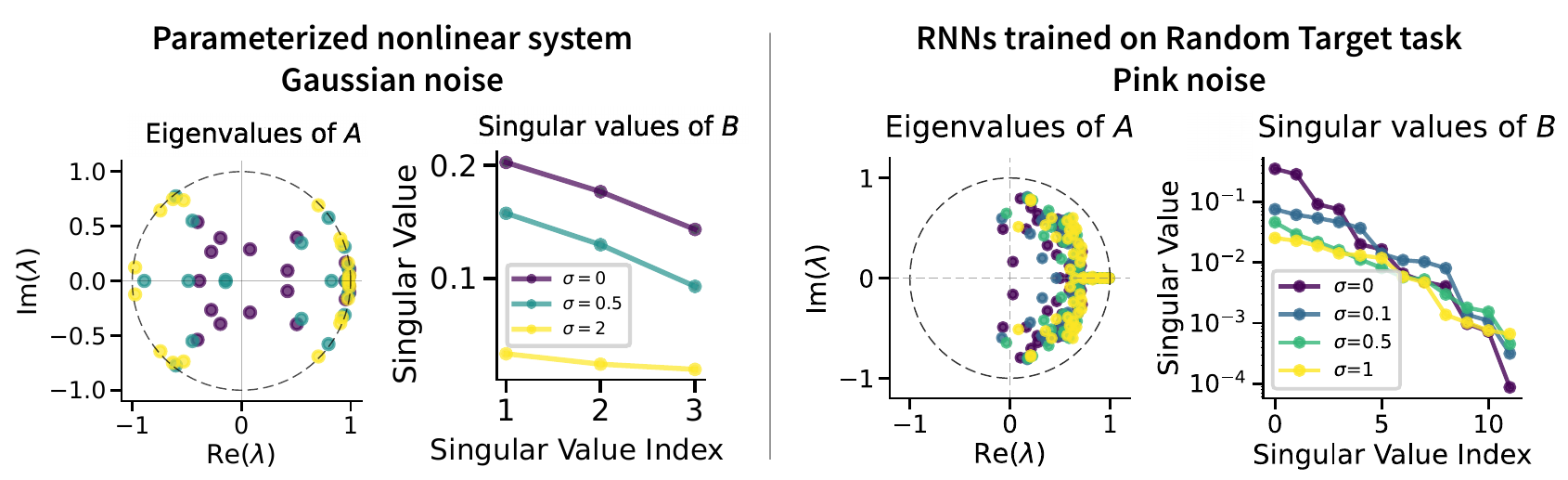}
    \caption{The eigenspectrum of the $A$ operator and singular spectrum of the $B$ operator when the input time-series is corrupted by Gaussian or pink noise of different variance.}
    \label{app:noise_effect}
\end{figure*}
Here, we also present a short theoretical discussion of this effect.

\label{app:omitted_variable_bias}
We consider the true system
\[
Y = AX + BU,
\]
where
\begin{itemize}
    \item $Y \in \mathbb{R}^{n \times T}$ are the next states,
    \item $X \in \mathbb{R}^{n \times T}$ are the current states,
    \item $U \in \mathbb{R}^{m \times T}$ are the inputs,
    \item $A \in \mathbb{R}^{n \times n}$ and $B \in \mathbb{R}^{n \times m}$.
\end{itemize}
We observe noisy inputs
\[
\tilde U = U + E,
\]
where $E$ is input noise. The regression becomes
\[
Y \approx \hat A X + \hat B \tilde U.
\]
We solve the regression problem with OLS by first stacking the regressors:
\[
Z = \begin{bmatrix} X \\ \tilde U \end{bmatrix}, 
\]
The OLS estimator is
\[
\begin{bmatrix} \hat A & \hat B \end{bmatrix}
= Y Z^\top (ZZ^\top)^{-1}.
\]
Expanding $Z$,
\[
ZZ^\top = \begin{bmatrix} X \\ \tilde U \end{bmatrix}\begin{bmatrix} X^\top & \tilde U^\top \end{bmatrix}=
\begin{bmatrix}
XX^\top & XU^\top + XE^\top \\
UX^\top + EX^\top & UU^\top + UE^\top + EU^\top + EE^\top
\end{bmatrix}.
\]
Assuming $E$ is zero-mean and independent,
\[
\mathbb{E}[ZZ^\top] =
\begin{bmatrix}
\Sigma_{xx} & \Sigma_{xu} \\
\Sigma_{ux} & \Sigma_{uu} + \Sigma_{ee}
\end{bmatrix},
\]
\[
\mathbb{E}[YZ^\top] =
\begin{bmatrix}
\Sigma_{xx} A^\top + \Sigma_{xu} B^\top \\
\Sigma_{ux} A^\top + \Sigma_{uu} B^\top
\end{bmatrix}.
\]
To gain intuition, we consider the scalar case where
\[
\sigma_{xx} = \mathrm{Var}(x), \quad
\sigma_{uu} = \mathrm{Var}(u), \quad
\sigma_{xu} = \mathrm{Cov}(x,u), \quad
\sigma_{ee} = \mathrm{Var}(E).
 \]
The least-squares estimates are
\[
\hat a = a + b \cdot \frac{\sigma_{xu}\,\sigma_{ee}}
{\sigma_{xx}(\sigma_{uu}+\sigma_{ee}) - \sigma_{xu}^2},
\]
\[
\hat b = b \cdot \frac{\sigma_{xx}\sigma_{uu} - \sigma_{xu}^2}
{\sigma_{xx}(\sigma_{uu}+\sigma_{ee}) - \sigma_{xu}^2}.
\]

We can see that for $\hat{b}$,  large $\sigma_{ee}$ in the denominator attenuates $\hat{b}$ toward zero. For $\hat{a}$, the direction of the bias is dependent on the signs of $b$, $\sigma_{xu}$ and relative weights of $\sigma_{xx}$ and $\sigma_{xu}$. In particular, when $b >0$ and $\sigma_{ee}$ is large, $\hat{a}$ can be inflated. Intuitively, when the state and input are strongly positively correlated and the input drives the state in the same direction, $\hat{a}$ can be overestimated to absorb the shared variance in the input. 

\section{Random Target Reach task}
\label{app:random_target}
\begin{figure*}[h] 
    \centering
    \includegraphics[width=\linewidth]{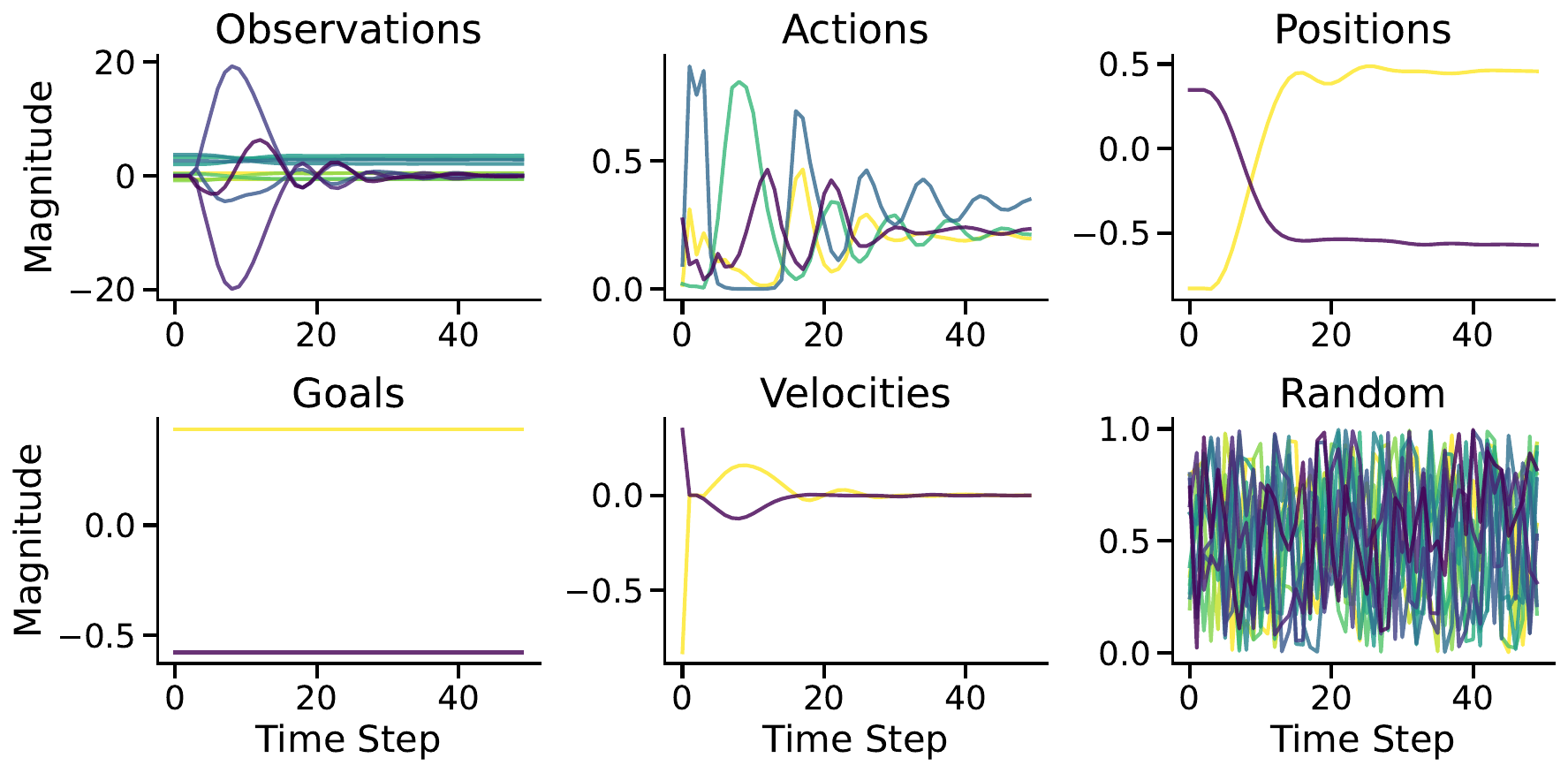}
    \caption{The true input (Observations) and different types of surrogate inputs during an example trial of the Random Target task.}
    \label{app:surrogate_inputs}
\end{figure*}

We trained recurrent neural network (RNN) policies to perform a random target reaching task in the MotorNet simulation environment \citep{Codol2024MotorNet}. We used code from \href{https://github.com/motornet-org/MotorNet}{https://github.com/motornet-org/MotorNet}. The effector was a \verb|ReluPointMass24| model, a 2D point-mass skeleton attached to 4 muscles and controlled by muscle activations. The environment provided a sequence of random goals and fingertip states. The objective of the policy was to minimize the distance between fingertip position and target over the course of each episode. At each time step, the model receives a 12-dimensional observation consisting of the proprioceptive input, visual input, and the last action taken by the model. The action space is 4-dimensional consisting of the activation of each muscle. Each network consisted of a single recurrent layer (64 hidden units) followed by a linear readout and sigmoid nonlinearity to produce bounded muscle activations. Training was carried out using the Adam optimizer with a learning rate of 0.001. The loss function was the mean L1 distance between fingertip trajectories and target trajectories across timesteps. We visualize the groundtruth input alongside different types of surrogate inputs during an example trial in Fig. \ref{app:surrogate_inputs}.

\section{Plume tracking task}

We used the plume tracking task implemented in \cite{singh2023emergent} and the training code in https://github.com/BruntonUWBio/plumetracknets. In short, the plume tracking environment is a 2D arena where an odor source emits puffs carried downwind by a steady flow. The wind can be constant, switch once, or switch multiple times during a trial. Each puff diffuses and drifts, producing intermittent odor encounters like in real plumes. The agent uses an actor–critic architecture with a vanilla RNN backbone, followed by separate two-layer MLPs for the actor and critic. At each timestep it receives three inputs: egocentric wind direction along the x-axis, wind direction along the y-axis, and local odor concentration. Based on its internal state, the actor outputs a two-dimensional action specifying turn rate and forward speed.

\paragraph{Hyperparameter sweep} To assess the optimal hyperparameters to use for InputDSA, we conducted a sweep of ranks and delays, as described by Sec. \ref{app: hpsweep}. By picking the minimum / elbow of the prediction error curves (AIC, MASE), we choose a delay of 40 and rank of 50 for all InputDSA computations on this dataset. 
\begin{figure*}[h] 
    \centering
    \includegraphics[width=\linewidth]{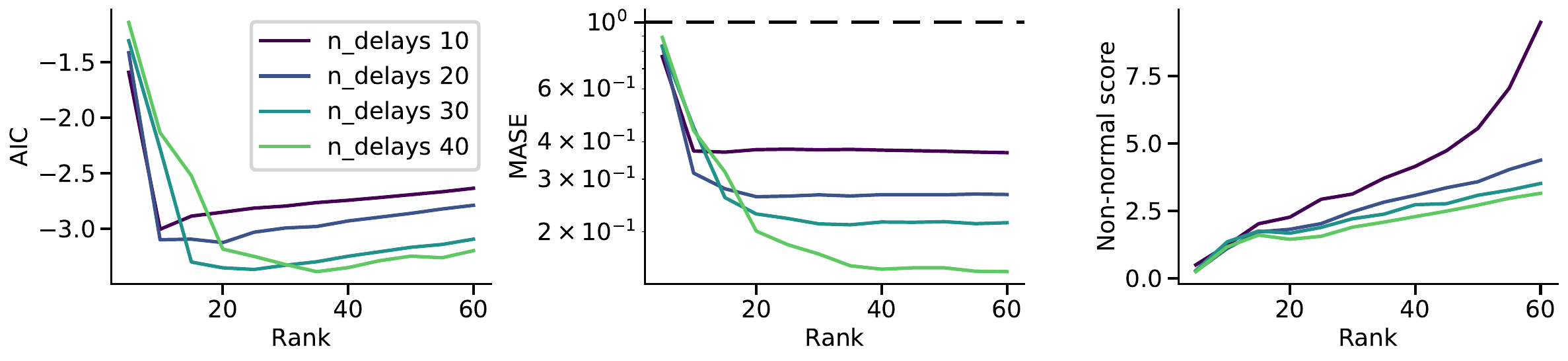}
    \caption{Hyperparameter sweep over number of delays and model rank on the plume tracking dataset.}
    \label{app:sweep_plume}
\end{figure*}
We also computed a non-normality score (the commutator score, $||AA^T - A^TA||_F^2$, which measures (as described) the degree to which a matrix is non-normal (with normality being defined as $A^TA = AA^T$). This measures the relevancy of non-normality in the prediction of the SubspaceDMDc model, which motivates the use of aligning the dynamics over orthogonal matrices, rather than invertible matrices -- although any invertible matrix is a coordinate transform, the dynamical system's transient before can change when the transform is non-orthogonal. Hence, capturing the full dynamical similarity of two systems can entail comparing up to orthogonal transform in these settings. Here, we find that a rank of 50 with a delay of 40 has a non-normality score close to 2.5, indicating that transient dynamics can be significantly effected. 

\section{Neural dataset}
\label{app:hyperparam_Luo}
The dataset published with \cite{luo2025transitions} can be found here: https://datadryad.org/dataset/doi:10.5061/dryad.sj3tx96dm. 

\paragraph{Hyperparameter sweep} We chose a delay of 5 for the delayed embedding and a rank of 6 for the reduced-rank regression on this dataset.

\begin{figure*}[h] 
    \centering
    \includegraphics[width=\linewidth]{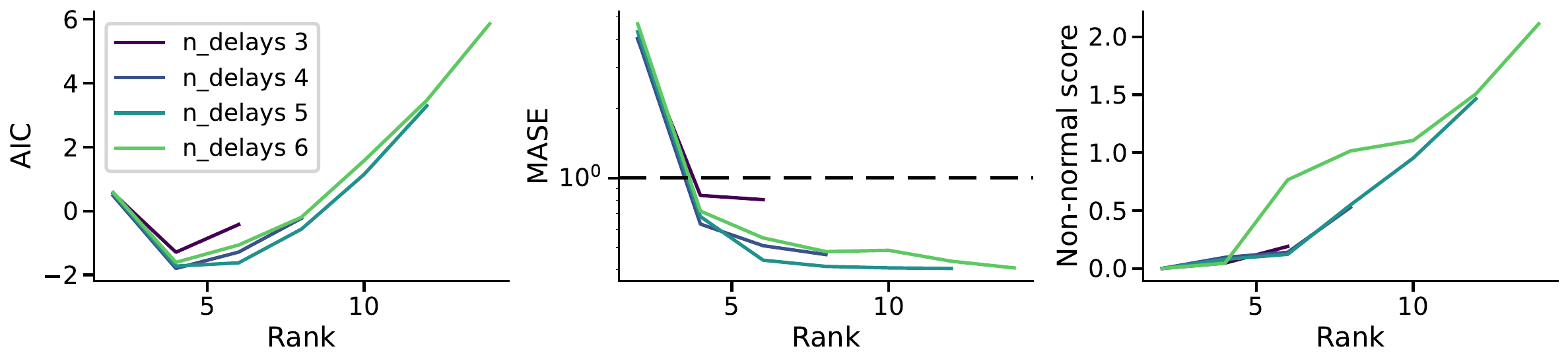}
    \caption{Hyperparameter sweep over number of delays and model rank on the processed Luo et al. (2025) dataset.}
    \label{app:sweep_Luo}
\end{figure*}

\subsection{Subspace Angle}

Given two dynamics matrices $A_x,A_y$, an orthonormal basis for each ($\tilde{A_x},\tilde{A_y}$) is first computed (for example, via SVD or QR decomposition). Then, the subspace angles are computed as:
\begin{align}
\quad \tilde{A}_x^T\tilde{A}_y &= U\Sigma V^T \\
    \theta_i &= \arccos(\sigma_i) 
\end{align}
Where $\sigma_i$ is the $i$-th singular value defined by $\Sigma$. $\theta_i$ is defined as the $i$ -th principal angle. We report the maximum principal angle between two dynamics operators, using the scipy.linalg.subspace\_angles function.

%% file: main.bbl
\begin{thebibliography}{93}
\providecommand{\natexlab}[1]{#1}
\providecommand{\url}[1]{\texttt{#1}}
\expandafter\ifx\csname urlstyle\endcsname\relax
  \providecommand{\doi}[1]{doi: #1}\else
  \providecommand{\doi}{doi: \begingroup \urlstyle{rm}\Url}\fi

\bibitem[Aldarondo et~al.(2024)Aldarondo, Merel, Marshall, Hasenclever,
  Klibaite, Gellis, Tassa, Wayne, Botvinick, and
  {\"O}lveczky]{aldarondo2024virtual}
Diego Aldarondo, Josh Merel, Jesse~D Marshall, Leonard Hasenclever, Ugne
  Klibaite, Amanda Gellis, Yuval Tassa, Greg Wayne, Matthew Botvinick, and
  Bence~P {\"O}lveczky.
\newblock A virtual rodent predicts the structure of neural activity across
  behaviours.
\newblock \emph{Nature}, 632\penalty0 (8025):\penalty0 594--602, 2024.

\bibitem[Arbabi \& Mezić(2017)Arbabi and Mezić]{arbabi_ergodic_2017}
Hassan Arbabi and Igor Mezić.
\newblock Ergodic {Theory}, {Dynamic} {Mode} {Decomposition}, and {Computation}
  of {Spectral} {Properties} of the {Koopman} {Operator}.
\newblock \emph{SIAM Journal on Applied Dynamical Systems}, \penalty0
  (4):\penalty0 2096--2126, 2017.

\bibitem[Asada \& Solano-Castellanos(2024)Asada and
  Solano-Castellanos]{asada2024controlcoherentkoopmanmodelingphysical}
H.~Harry Asada and Jose~A. Solano-Castellanos.
\newblock Control-coherent koopman modeling: A physical modeling approach,
  2024.
\newblock URL \url{https://arxiv.org/abs/2403.16306}.

\bibitem[Brunton et~al.(2017)Brunton, Brunton, Proctor, Kaiser, and
  Kutz]{brunton_chaos_2017}
Steven~L. Brunton, Bingni~W. Brunton, Joshua~L. Proctor, Eurika Kaiser, and
  J.~Nathan Kutz.
\newblock Chaos as an {Intermittently} {Forced} {Linear} {System}.
\newblock \emph{Nature Communications}, 8\penalty0 (1):\penalty0 19, December
  2017.
\newblock ISSN 2041-1723.
\newblock \doi{10.1038/s41467-017-00030-8}.
\newblock URL \url{http://arxiv.org/abs/1608.05306}.
\newblock arXiv:1608.05306 [nlin].

\bibitem[Budišić et~al.(2012)Budišić, Mohr, and
  Mezić]{budisic_applied_2012}
Marko Budišić, Ryan Mohr, and Igor Mezić.
\newblock Applied {Koopmanism}.
\newblock \emph{Chaos: An Interdisciplinary Journal of Nonlinear Science},
  22\penalty0 (4):\penalty0 047510, December 2012.

\bibitem[Burak \& Fiete(2009{\natexlab{a}})Burak and Fiete]{burak2009accurate}
Yoram Burak and Ila~R Fiete.
\newblock Accurate path integration in continuous attractor network models of
  grid cells.
\newblock \emph{PLoS computational biology}, 5\penalty0 (2):\penalty0 e1000291,
  2009{\natexlab{a}}.

\bibitem[Burak \& Fiete(2009{\natexlab{b}})Burak and
  Fiete]{burak_accurate_2009}
Yoram Burak and Ila~R Fiete.
\newblock Accurate path integration in continuous attractor network models of
  grid cells.
\newblock \emph{PLoS Comput. Biol.}, 5\penalty0 (2):\penalty0 e1000291,
  February 2009{\natexlab{b}}.

\bibitem[Chaudhuri et~al.(2019)Chaudhuri, Ger{\c{c}}ek, Pandey, Peyrache, and
  Fiete]{chaudhuri2019intrinsic}
Rishidev Chaudhuri, Berk Ger{\c{c}}ek, Biraj Pandey, Adrien Peyrache, and Ila
  Fiete.
\newblock The intrinsic attractor manifold and population dynamics of a
  canonical cognitive circuit across waking and sleep.
\newblock \emph{Nature neuroscience}, 22\penalty0 (9):\penalty0 1512--1520,
  2019.

\bibitem[Chen et~al.(2024)Chen, Vedovati, Braver, and
  Ching]{ChenVedovatiBraverChing2024DFORM}
Ruiqi Chen, Giacomo Vedovati, Todd Braver, and ShiNung Ching.
\newblock Dform: Diffeomorphic vector field alignment for assessing dynamics
  across learned models.
\newblock \emph{arXiv preprint arXiv:2402.09735}, 2024.
\newblock \doi{10.48550/arXiv.2402.09735}.

\bibitem[Churchland et~al.(2012)Churchland, Cunningham, Kaufman, Foster,
  Nuyujukian, Ryu, and Shenoy]{Churchland2012Nature}
Mark~M. Churchland, John~P. Cunningham, Matthew~T. Kaufman, Justin~D. Foster,
  Paul Nuyujukian, Stephen~I. Ryu, and Krishna~V. Shenoy.
\newblock Neural population dynamics during reaching.
\newblock \emph{Nature}, 487\penalty0 (7405):\penalty0 51--56, 2012.

\bibitem[Codol et~al.(2024{\natexlab{a}})Codol, Krishna, Lajoie, and
  Perich]{codol2024brain}
Olivier Codol, Nanda~H Krishna, Guillaume Lajoie, and Matthew~G Perich.
\newblock Brain-like neural dynamics for behavioral control develop through
  reinforcement learning.
\newblock \emph{bioRxiv}, pp.\  2024--10, 2024{\natexlab{a}}.

\bibitem[Codol et~al.(2024{\natexlab{b}})Codol, Michaels, Kashefi, Pruszynski,
  and Gribble]{Codol2024MotorNet}
Olivier Codol, Jonathan~A. Michaels, Mehrdad Kashefi, J.~Andrew Pruszynski, and
  Paul~L. Gribble.
\newblock Motornet: A python toolbox for controlling differentiable
  biomechanical effectors with artificial neural networks.
\newblock \emph{eLife}, 13:\penalty0 RP88591, 2024{\natexlab{b}}.
\newblock \doi{10.7554/eLife.88591}.

\bibitem[Colbrook et~al.(2023)Colbrook, Ayton, and Szőke]{Colbrook_2023}
Matthew~J. Colbrook, Lorna~J. Ayton, and Máté Szőke.
\newblock Residual dynamic mode decomposition: robust and verified koopmanism.
\newblock \emph{Journal of Fluid Mechanics}, 955, 2023.
\newblock ISSN 1469-7645.
\newblock \doi{10.1017/jfm.2022.1052}.

\bibitem[Compte et~al.(2000)Compte, Brunel, Goldman-Rakic, and
  Wang]{Compte2000SpatialWM}
Albert Compte, Nicolas Brunel, Patricia~S. Goldman-Rakic, and Xiao-Jing Wang.
\newblock Synaptic mechanisms and network dynamics underlying spatial working
  memory in a cortical network model.
\newblock \emph{Cerebral Cortex}, 10\penalty0 (9):\penalty0 910--923, 2000.
\newblock \doi{10.1093/cercor/10.9.910}.

\bibitem[Cotler et~al.(2023)Cotler, Tai, Hernández, Elias, and
  Sussillo]{cotler2023analyzingpopulationsneuralnetworks}
Jordan Cotler, Kai~Sheng Tai, Felipe Hernández, Blake Elias, and David
  Sussillo.
\newblock Analyzing populations of neural networks via dynamical model
  embedding, 2023.
\newblock URL \url{https://arxiv.org/abs/2302.14078}.

\bibitem[Eisen et~al.(2025)Eisen, Ostrow, Chandra, Kozachkov, Miller, and
  Fiete]{eisen2025characterizing}
Adam~J Eisen, Mitchell Ostrow, Sarthak Chandra, Leo Kozachkov, Earl~K Miller,
  and Ila~R Fiete.
\newblock Characterizing control between interacting subsystems with deep
  jacobian estimation.
\newblock \emph{arXiv preprint arXiv:2507.01946}, 2025.

\bibitem[Feigenbaum(1978)]{feigenbaum1978quantitative}
Mitchell~J Feigenbaum.
\newblock Quantitative universality for a class of nonlinear transformations.
\newblock \emph{Journal of statistical physics}, 19\penalty0 (1):\penalty0
  25--52, 1978.

\bibitem[Flint et~al.(2012)Flint, Lindberg, Jordan, Miller, and
  Slutzky]{Flint2012FieldPotentials}
Robert~D. Flint, Eric~W. Lindberg, Luke~R. Jordan, Lee~E. Miller, and Marc~W.
  Slutzky.
\newblock Accurate decoding of reaching movements from field potentials in the
  absence of spikes.
\newblock \emph{Journal of Neural Engineering}, 9\penalty0 (4):\penalty0
  046006, 2012.

\bibitem[Funahashi et~al.(1989)Funahashi, Bruce, and
  Goldman-Rakic]{Funahashi1989MnemonicCoding}
Shintaro Funahashi, Charles~J. Bruce, and Patricia~S. Goldman-Rakic.
\newblock Mnemonic coding of visual space in the monkey's dorsolateral
  prefrontal cortex.
\newblock \emph{Journal of Neurophysiology}, 61\penalty0 (2):\penalty0
  331--349, 1989.
\newblock \doi{10.1152/jn.1989.61.2.331}.

\bibitem[Fuster \& Alexander(1971)Fuster and
  Alexander]{FusterAlexander1971ShortTermMemory}
Joaquin~M. Fuster and Gary~E. Alexander.
\newblock Neuron activity related to short-term memory.
\newblock \emph{Science}, 173\penalty0 (3997):\penalty0 652--654, 1971.
\newblock \doi{10.1126/science.173.3997.652}.

\bibitem[Galgali et~al.(2023)Galgali, Sahani, and Mante]{galgali_residual_2023}
Aniruddh~R. Galgali, Maneesh Sahani, and Valerio Mante.
\newblock Residual dynamics resolves recurrent contributions to neural
  computation.
\newblock \emph{Nature Neuroscience}, 2023.

\bibitem[Gallego et~al.(2018)Gallego, Perich, Naufel, Ethier, Solla, and
  Miller]{gallego_cortical_2018}
Juan~A. Gallego, Matthew~G. Perich, Stephanie~N. Naufel, Christian Ethier,
  Sara~A. Solla, and Lee~E. Miller.
\newblock Cortical population activity within a preserved neural manifold
  underlies multiple motor behaviors.
\newblock \emph{Nature Communications}, 9:\penalty0 4233, October 2018.
\newblock ISSN 2041-1723.
\newblock \doi{10.1038/s41467-018-06560-z}.
\newblock URL \url{https://www.ncbi.nlm.nih.gov/pmc/articles/PMC6185944/}.

\bibitem[Gardner et~al.(2022)Gardner, Hermansen, Pachitariu, Burak, Baas, Dunn,
  Moser, and Moser]{gardner2022toroidal}
Richard~J Gardner, Erik Hermansen, Marius Pachitariu, Yoram Burak, Nils~A Baas,
  Benjamin~A Dunn, May-Britt Moser, and Edvard~I Moser.
\newblock Toroidal topology of population activity in grid cells.
\newblock \emph{Nature}, 602\penalty0 (7895):\penalty0 123--128, 2022.

\bibitem[Goldman-Rakic(1995)]{GoldmanRakic1995CellularBasisWM}
Patricia~S. Goldman-Rakic.
\newblock Cellular basis of working memory.
\newblock \emph{Neuron}, 14\penalty0 (3):\penalty0 477--485, 1995.
\newblock \doi{10.1016/0896-6273(95)90304-6}.

\bibitem[Gosztolai et~al.(2025)Gosztolai, Peach, Arnaudon, Barahona, and
  Vandergheynst]{gosztolai_marble_2025}
Adam Gosztolai, Robert~L. Peach, Alexis Arnaudon, Mauricio Barahona, and Pierre
  Vandergheynst.
\newblock {MARBLE}: interpretable representations of neural population dynamics
  using geometric deep learning.
\newblock \emph{Nature Methods}, 22\penalty0 (3):\penalty0 612--620, March
  2025.
\newblock ISSN 1548-7105.
\newblock \doi{10.1038/s41592-024-02582-2}.
\newblock URL \url{https://doi.org/10.1038/s41592-024-02582-2}.

\bibitem[Grillner(2006)]{Grillner2006NetworksInMotion}
Sten Grillner.
\newblock Biological pattern generation: The cellular and computational logic
  of networks in motion.
\newblock \emph{Neuron}, 52\penalty0 (5):\penalty0 751--766, 2006.
\newblock \doi{10.1016/j.neuron.2006.11.008}.

\bibitem[Guilhot et~al.(2024)Guilhot, W{\'o}jcik, Achterberg, and
  Costa]{guilhot2024dynamical}
Quentin Guilhot, Micha{\l}~J W{\'o}jcik, Jascha Achterberg, and Rui~Ponte
  Costa.
\newblock Dynamical similarity analysis uniquely captures how computations
  develop in rnns.
\newblock 2024.

\bibitem[Haseli et~al.(2025)Haseli, Mezić, and
  Cortés]{haseli2025roadskoopmanoperatortheory}
Masih Haseli, Igor Mezić, and Jorge Cortés.
\newblock Two roads to koopman operator theory for control: Infinite input
  sequences and operator families, 2025.
\newblock URL \url{https://arxiv.org/abs/2510.15166}.

\bibitem[Hatsopoulos et~al.(2007)Hatsopoulos, Xu, and
  Amit]{Hatsopoulos2007Fragments}
Nicholas~G. Hatsopoulos, Qingqing Xu, and Yali Amit.
\newblock Encoding of movement fragments in the motor cortex.
\newblock \emph{Journal of Neuroscience}, 27\penalty0 (19):\penalty0
  5105--5114, May 2007.
\newblock \doi{10.1523/JNEUROSCI.3570-06.2007}.
\newblock URL \url{https://www.jneurosci.org/content/27/19/5105}.

\bibitem[Hohenberg \& Halperin(1977)Hohenberg and Halperin]{hohenbergcritical}
P.~C. Hohenberg and B.~I. Halperin.
\newblock Theory of dynamic critical phenomena.
\newblock \emph{Rev. Mod. Phys.}, 49:\penalty0 435--479, Jul 1977.
\newblock \doi{10.1103/RevModPhys.49.435}.
\newblock URL \url{https://link.aps.org/doi/10.1103/RevModPhys.49.435}.

\bibitem[Huang et~al.(2025)Huang, Singh, Martinelli, and
  Rajan]{huang2025measuring}
Ann Huang, Satpreet~H Singh, Flavio Martinelli, and Kanaka Rajan.
\newblock Measuring and controlling solution degeneracy across task-trained
  recurrent neural networks.
\newblock \emph{ArXiv}, pp.\  arXiv--2410, 2025.

\bibitem[Huh et~al.(2024)Huh, Cheung, Wang, and
  Isola]{huh2024platonicrepresentationhypothesis}
Minyoung Huh, Brian Cheung, Tongzhou Wang, and Phillip Isola.
\newblock The platonic representation hypothesis, 2024.
\newblock URL \url{https://arxiv.org/abs/2405.07987}.

\bibitem[Ichinaga et~al.(2024)Ichinaga, Andreuzzi, Demo, Tezzele, Lapo, Rozza,
  Brunton, and Kutz]{ichinaga_pydmd_2024}
Sara~M. Ichinaga, Francesco Andreuzzi, Nicola Demo, Marco Tezzele, Karl Lapo,
  Gianluigi Rozza, Steven~L. Brunton, and J.~Nathan Kutz.
\newblock {PyDMD}: {A} {Python} package for robust dynamic mode decomposition,
  February 2024.
\newblock URL \url{http://arxiv.org/abs/2402.07463}.
\newblock arXiv:2402.07463 [stat].

\bibitem[Juang \& Pappa(1985)Juang and Pappa]{juang_eigensystem_1985}
J.~N. Juang and R.~S. Pappa.
\newblock An {Eigensystem} {Realization} {Algorithm} ({ERA}) for modal
  parameter identification and model reduction.
\newblock April 1985.
\newblock URL \url{https://ntrs.nasa.gov/citations/19850022899}.

\bibitem[Kao \& Hennequin(2019)Kao and Hennequin]{kao2019neuroscience}
Ta-Chu Kao and Guillaume Hennequin.
\newblock Neuroscience out of control: control-theoretic perspectives on neural
  circuit dynamics.
\newblock \emph{Current opinion in neurobiology}, 58:\penalty0 122--129, 2019.

\bibitem[Kennel et~al.(1992)Kennel, Brown, and Abarbanel]{kennelfalseneighbors}
Matthew~B. Kennel, Reggie Brown, and Henry D.~I. Abarbanel.
\newblock Determining embedding dimension for phase-space reconstruction using
  a geometrical construction.
\newblock \emph{Phys. Rev. A}, 45:\penalty0 3403--3411, Mar 1992.
\newblock \doi{10.1103/PhysRevA.45.3403}.
\newblock URL \url{https://link.aps.org/doi/10.1103/PhysRevA.45.3403}.

\bibitem[Kiehn(2016)]{Kiehn2016SpinalCircuits}
Ole Kiehn.
\newblock Decoding the organization of spinal circuits that control locomotion.
\newblock \emph{Nature Reviews Neuroscience}, 17\penalty0 (4):\penalty0
  224--238, 2016.
\newblock \doi{10.1038/nrn.2016.9}.

\bibitem[Koopman(1931)]{Koopman1931}
B.~O. Koopman.
\newblock Hamiltonian systems and transformation in hilbert space.
\newblock \emph{Proceedings of the National Academy of Sciences}, 17\penalty0
  (5):\penalty0 315--318, 1931.
\newblock \doi{10.1073/pnas.17.5.315}.
\newblock URL \url{https://www.pnas.org/doi/abs/10.1073/pnas.17.5.315}.

\bibitem[Korda \& Mezić(2018)Korda and Mezić]{mezickoopmancontrol1}
Milan Korda and Igor Mezić.
\newblock Linear predictors for nonlinear dynamical systems: Koopman operator
  meets model predictive control.
\newblock \emph{Automatica}, 93:\penalty0 149--160, 2018.
\newblock ISSN 0005-1098.
\newblock \doi{https://doi.org/10.1016/j.automatica.2018.03.046}.
\newblock URL
  \url{https://www.sciencedirect.com/science/article/pii/S000510981830133X}.

\bibitem[Kornblith et~al.(2019)Kornblith, Norouzi, Lee, and
  Hinton]{kornblith_similarity_2019}
Simon Kornblith, Mohammad Norouzi, Honglak Lee, and Geoffrey Hinton.
\newblock Similarity of {Neural} {Network} {Representations} {Revisited}, July
  2019.
\newblock URL \url{http://arxiv.org/abs/1905.00414}.
\newblock arXiv:1905.00414 [cs, q-bio, stat].

\bibitem[Kriegeskorte et~al.(2008)Kriegeskorte, Mur, and
  Bandettini]{Kriegeskorte2008RSA}
Nikolaus Kriegeskorte, Marieke Mur, and Peter Bandettini.
\newblock Representational similarity analysis — connecting the branches of
  systems neuroscience.
\newblock \emph{Frontiers in Systems Neuroscience}, 2:\penalty0 4, 2008.
\newblock \doi{10.3389/neuro.06.004.2008}.

\bibitem[Lazzari \& Saxena(2025)Lazzari and Saxena]{lazzari2025multitasking}
John Lazzari and Shreya Saxena.
\newblock Multitasking recurrent networks utilize compositional strategies for
  control of movement.
\newblock \emph{bioRxiv}, pp.\  2025--09, 2025.

\bibitem[Lin \& Kriegeskorte(2024)Lin and Kriegeskorte]{Trsa}
Baihan Lin and Nikolaus Kriegeskorte.
\newblock The topology and geometry of neural representations.
\newblock \emph{Proceedings of the National Academy of Sciences}, 121\penalty0
  (42):\penalty0 e2317881121, 2024.
\newblock \doi{10.1073/pnas.2317881121}.
\newblock URL \url{https://www.pnas.org/doi/abs/10.1073/pnas.2317881121}.

\bibitem[Logiaco et~al.(2021)Logiaco, Abbott, and Escola]{logiaco2021thalamic}
Laureline Logiaco, LF~Abbott, and Sean Escola.
\newblock Thalamic control of cortical dynamics in a model of flexible motor
  sequencing.
\newblock \emph{Cell reports}, 35\penalty0 (9), 2021.

\bibitem[Luenberger(1979)]{luenberger1979dynamic}
David~G Luenberger.
\newblock Dynamic systems, 1979.

\bibitem[Luo et~al.(2025)Luo, Kim, Gupta, Bondy, Kopec, Elliott, DePasquale,
  and Brody]{luo2025transitions}
Thomas~Zhihao Luo, Timothy~Doyeon Kim, Diksha Gupta, Adrian~G Bondy, Charles~D
  Kopec, Verity~A Elliott, Brian DePasquale, and Carlos~D Brody.
\newblock Transitions in dynamical regime and neural mode during perceptual
  decisions.
\newblock \emph{Nature}, pp.\  1--11, 2025.

\bibitem[Lusch et~al.(2018)Lusch, Kutz, and Brunton]{lusch_deep_2018}
Bethany Lusch, J.~Nathan Kutz, and Steven~L. Brunton.
\newblock Deep learning for universal linear embeddings of nonlinear dynamics.
\newblock \emph{Nature Communications}, 9\penalty0 (1):\penalty0 4950, November
  2018.
\newblock ISSN 2041-1723.
\newblock \doi{10.1038/s41467-018-07210-0}.
\newblock URL \url{https://www.nature.com/articles/s41467-018-07210-0}.

\bibitem[Madhav \& Cowan(2020)Madhav and Cowan]{madhav2020synergy}
Manu~S Madhav and Noah~J Cowan.
\newblock The synergy between neuroscience and control theory: the nervous
  system as inspiration for hard control challenges.
\newblock \emph{Annual Review of Control, Robotics, and Autonomous Systems},
  3\penalty0 (1):\penalty0 243--267, 2020.

\bibitem[Maheswaranathan et~al.(2019)Maheswaranathan, Williams, Golub, Ganguli,
  and Sussillo]{maheswaranathan2019universality}
Niru Maheswaranathan, Alex Williams, Matthew Golub, Surya Ganguli, and David
  Sussillo.
\newblock Universality and individuality in neural dynamics across large
  populations of recurrent networks.
\newblock In \emph{Advances in neural information processing systems}, pp.\
  15629--15641, 2019.

\bibitem[Mante et~al.(2013{\natexlab{a}})Mante, Sussillo, Shenoy, and
  Newsome]{mante2013context}
Valerio Mante, David Sussillo, Krishna~V Shenoy, and William~T Newsome.
\newblock Context-dependent computation by recurrent dynamics in prefrontal
  cortex.
\newblock \emph{nature}, 503\penalty0 (7474):\penalty0 78--84,
  2013{\natexlab{a}}.

\bibitem[Mante et~al.(2013{\natexlab{b}})Mante, Sussillo, Shenoy, and
  Newsome]{mante_context-dependent_2013}
Valerio Mante, David Sussillo, Krishna~V. Shenoy, and William~T. Newsome.
\newblock Context-dependent computation by recurrent dynamics in prefrontal
  cortex.
\newblock \emph{Nature}, 503\penalty0 (7474):\penalty0 78--84, November
  2013{\natexlab{b}}.
\newblock ISSN 1476-4687.
\newblock \doi{10.1038/nature12742}.
\newblock URL \url{https://www.nature.com/articles/nature12742}.
\newblock Number: 7474 Publisher: Nature Publishing Group.

\bibitem[Marder \& Bucher(2001)Marder and Bucher]{MarderBucher2001CPG}
Eve Marder and Dirk Bucher.
\newblock Central pattern generators and the control of rhythmic movements.
\newblock \emph{Current Biology}, 11\penalty0 (23):\penalty0 R986--R996, 2001.
\newblock \doi{10.1016/S0960-9822(01)00581-4}.

\bibitem[Mezic(2016)]{mezic_comparison_2016}
Igor Mezic.
\newblock On {Comparison} of {Dynamics} of {Dissipative} and {Finite}-{Time}
  {Systems} {Using} {Koopman} {Operator} {Methods}**{The} funding provided by
  {ARO} {Grant} {W911NF}-11-1-0511.
\newblock \emph{IFAC-PapersOnLine}, 49:\penalty0 454--461, December 2016.
\newblock \doi{10.1016/j.ifacol.2016.10.207}.

\bibitem[Mezić \& Banaszuk(2004)Mezić and Banaszuk]{mezic_comparison_2004}
Igor Mezić and Andrzej Banaszuk.
\newblock Comparison of systems with complex behavior.
\newblock \emph{Physica D: Nonlinear Phenomena}, 197\penalty0 (1):\penalty0
  101--133, October 2004.
\newblock ISSN 0167-2789.
\newblock \doi{10.1016/j.physd.2004.06.015}.
\newblock URL
  \url{https://www.sciencedirect.com/science/article/pii/S0167278904002507}.

\bibitem[Nair et~al.(2023)Nair, Karigo, Yang, Ganguli, Schnitzer, Linderman,
  Anderson, and Kennedy]{nair2023approximate}
Aditya Nair, Tomomi Karigo, Bin Yang, Surya Ganguli, Mark~J Schnitzer, Scott~W
  Linderman, David~J Anderson, and Ann Kennedy.
\newblock An approximate line attractor in the hypothalamus encodes an
  aggressive state.
\newblock \emph{Cell}, 186\penalty0 (1):\penalty0 178--193, 2023.

\bibitem[Nejatbakhsh et~al.(2024)Nejatbakhsh, Geadah, Williams, and
  Lipshutz]{nejatbakhsh2024comparing}
Amin Nejatbakhsh, Victor Geadah, Alex~H Williams, and David Lipshutz.
\newblock Comparing noisy neural population dynamics using optimal transport
  distances.
\newblock \emph{arXiv preprint arXiv:2412.14421}, 2024.

\bibitem[Ostrow et~al.(2023)Ostrow, Eisen, Kozachkov, and
  Fiete]{ostrow2024beyond}
Mitchell Ostrow, Adam Eisen, Leo Kozachkov, and Ila Fiete.
\newblock Beyond geometry: Comparing the temporal structure of computation in
  neural circuits with dynamical similarity analysis.
\newblock \emph{Advances in Neural Information Processing Systems}, 36, 2023.

\bibitem[Ostrow et~al.(2024)Ostrow, Eisen, and Fiete]{ostrow2024delay}
Mitchell Ostrow, Adam Eisen, and Ila Fiete.
\newblock Delay embedding theory of neural sequence models.
\newblock \emph{arXiv preprint arXiv:2406.11993}, 2024.
\newblock URL \url{https://arxiv.org/abs/2406.11993}.

\bibitem[Overschee \& Moor(1994)Overschee and Moor]{n4sid}
Peter~Van Overschee and Bart~De Moor.
\newblock N4sid: Subspace algorithms for the identification of combined
  deterministic-stochastic systems.
\newblock \emph{Automatica}, 30:\penalty0 75--93, 1994.
\newblock URL \url{https://api.semanticscholar.org/CorpusID:28586805}.

\bibitem[O’Shea et~al.(2022)O’Shea, Duncker, Goo, Sun, Vyas, Trautmann,
  Diester, Ramakrishnan, Deisseroth, Sahani, et~al.]{o2022direct}
Daniel~J O’Shea, Lea Duncker, Werapong Goo, Xulu Sun, Saurabh Vyas, Eric~M
  Trautmann, Ilka Diester, Charu Ramakrishnan, Karl Deisseroth, Maneesh Sahani,
  et~al.
\newblock Direct neural perturbations reveal a dynamical mechanism for robust
  computation.
\newblock \emph{bioRxiv}, pp.\  2022--12, 2022.

\bibitem[Perich et~al.(2020)Perich, Arlt, Soares, Young, Mosher, Minxha,
  Carter, Rutishauser, Rudebeck, Harvey, et~al.]{perich2020inferring}
Matthew~G Perich, Charlotte Arlt, Sofia Soares, Megan~E Young, Clayton~P
  Mosher, Juri Minxha, Eugene Carter, Ueli Rutishauser, Peter~H Rudebeck,
  Christopher~D Harvey, et~al.
\newblock Inferring brain-wide interactions using data-constrained recurrent
  neural network models.
\newblock \emph{BioRxiv}, pp.\  2020--12, 2020.

\bibitem[Prinz et~al.(2004)Prinz, Bucher, and Marder]{prinz_similar_2004}
Astrid~A Prinz, Dirk Bucher, and Eve Marder.
\newblock Similar network activity from disparate circuit parameters.
\newblock \emph{Nature Neuroscience}, 7\penalty0 (12):\penalty0 1345--1352,
  December 2004.
\newblock ISSN 1546-1726.
\newblock \doi{10.1038/nn1352}.
\newblock URL \url{https://doi.org/10.1038/nn1352}.

\bibitem[Proctor et~al.(2016{\natexlab{a}})Proctor, Brunton, and
  Kutz]{proctor2016dynamic}
Joshua~L Proctor, Steven~L Brunton, and J~Nathan Kutz.
\newblock Dynamic mode decomposition with control.
\newblock \emph{SIAM Journal on Applied Dynamical Systems}, 15\penalty0
  (1):\penalty0 142--161, 2016{\natexlab{a}}.

\bibitem[Proctor et~al.(2016{\natexlab{b}})Proctor, Brunton, and
  Kutz]{proctor_generalizing_2016}
Joshua~L. Proctor, Steven~L. Brunton, and J.~Nathan Kutz.
\newblock Generalizing {Koopman} {Theory} to allow for inputs and control,
  February 2016{\natexlab{b}}.
\newblock URL \url{http://arxiv.org/abs/1602.07647}.
\newblock arXiv:1602.07647 [math].

\bibitem[Raghu et~al.(2017)Raghu, Gilmer, Yosinski, and
  Sohl-Dickstein]{raghu_svcca_2017}
Maithra Raghu, Justin Gilmer, Jason Yosinski, and Jascha Sohl-Dickstein.
\newblock {SVCCA}: {Singular} {Vector} {Canonical} {Correlation} {Analysis} for
  {Deep} {Learning} {Dynamics} and {Interpretability}, November 2017.
\newblock URL \url{http://arxiv.org/abs/1706.05806}.
\newblock arXiv:1706.05806 [cs, stat].

\bibitem[Rajan et~al.(2010)Rajan, Abbott, and Sompolinsky]{RajanEtAl2010}
Kanaka Rajan, L.~F. Abbott, and Haim Sompolinsky.
\newblock Stimulus-dependent suppression of chaos in recurrent neural networks.
\newblock \emph{Physical Review E}, 82\penalty0 (1):\penalty0 011903, July
  2010.
\newblock \doi{10.1103/PhysRevE.82.011903}.

\bibitem[Redman et~al.(2024{\natexlab{a}})Redman, Acosta, Acosta-Mendoza, and
  Miolane]{redman2024not}
William Redman, Francisco Acosta, Santiago Acosta-Mendoza, and Nina Miolane.
\newblock Not so griddy: Internal representations of rnns path integrating more
  than one agent.
\newblock \emph{Advances in Neural Information Processing Systems},
  37:\penalty0 22657--22689, 2024{\natexlab{a}}.

\bibitem[Redman et~al.(2024{\natexlab{b}})Redman, Bello-Rivas, Fonoberova,
  Mohr, Kevrekidis, and Mezi\'{c}]{redman2024}
William~T. Redman, Juan Bello-Rivas, Maria Fonoberova, Ryan Mohr, Yannis~G.
  Kevrekidis, and Igor Mezi\'{c}.
\newblock Identifying equivalent training dynamics.
\newblock In A.~Globerson, L.~Mackey, D.~Belgrave, A.~Fan, U.~Paquet,
  J.~Tomczak, and C.~Zhang (eds.), \emph{Advances in Neural Information
  Processing Systems}, volume~37, pp.\  23603--23629. Curran Associates, Inc.,
  2024{\natexlab{b}}.
\newblock URL
  \url{https://proceedings.neurips.cc/paper_files/paper/2024/file/2a07348a6a7b2c208ab5cb1ee0e78ab5-Paper-Conference.pdf}.

\bibitem[Sani et~al.(2024)Sani, Pesaran, and Shanechi]{sani2024dissociative}
Omid~G. Sani, Bijan Pesaran, and Maryam~M. Shanechi.
\newblock Dissociative and prioritized modeling of behaviorally relevant neural
  dynamics using recurrent neural networks.
\newblock \emph{Nature Neuroscience}, 27\penalty0 (10):\penalty0 2033--2045,
  2024.
\newblock \doi{10.1038/s41593-024-01731-2}.
\newblock URL \url{https://www.nature.com/articles/s41593-024-01731-2}.

\bibitem[Schaeffer et~al.(2020)Schaeffer, Khona, Meshulam, International, and
  Fiete]{schaeffer_reverse-engineering_2020}
Rylan Schaeffer, Mikail Khona, Leenoy Meshulam, Brain~Laboratory International,
  and Ila Fiete.
\newblock Reverse-engineering recurrent neural network solutions to a
  hierarchical inference task for mice.
\newblock In \emph{Advances in {Neural} {Information} {Processing} {Systems}},
  volume~33, pp.\  4584--4596. Curran Associates, Inc., 2020.
\newblock URL
  \url{https://papers.nips.cc/paper/2020/hash/30f0641c041f03d94e95a76b9d8bd58f-Abstract.html}.

\bibitem[Schmid(2010)]{schmid_dynamic_2010}
Peter~J. Schmid.
\newblock Dynamic mode decomposition of numerical and experimental data.
\newblock \emph{Journal of Fluid Mechanics}, 656:\penalty0 5--28, August 2010.
\newblock ISSN 0022-1120, 1469-7645.
\newblock \doi{10.1017/S0022112010001217}.
\newblock URL
  \url{https://www.cambridge.org/core/product/identifier/S0022112010001217/type/journal_article}.

\bibitem[Schmid(2022)]{schmid2022dynamic}
Peter~J Schmid.
\newblock Dynamic mode decomposition and its variants.
\newblock \emph{Annual Review of Fluid Mechanics}, 54\penalty0 (1):\penalty0
  225--254, 2022.

\bibitem[Schrimpf et~al.(2018)Schrimpf, Kubilius, Hong, Majaj, Rajalingham,
  Issa, Kar, Bashivan, Prescott-Roy, Geiger, Schmidt, Yamins, and
  DiCarlo]{schrimpf_brain-score_2018}
Martin Schrimpf, Jonas Kubilius, Ha~Hong, Najib~J. Majaj, Rishi Rajalingham,
  Elias~B. Issa, Kohitij Kar, Pouya Bashivan, Jonathan Prescott-Roy, Franziska
  Geiger, Kailyn Schmidt, Daniel L.~K. Yamins, and James~J. DiCarlo.
\newblock Brain-{Score}: {Which} {Artificial} {Neural} {Network} for {Object}
  {Recognition} is most {Brain}-{Like}?
\newblock preprint, Neuroscience, September 2018.
\newblock URL \url{http://biorxiv.org/lookup/doi/10.1101/407007}.

\bibitem[Shine et~al.(2021)Shine, M{\"u}ller, Munn, Cabral, Moran, and
  Breakspear]{shine2021computational}
James~M Shine, Eli~J M{\"u}ller, Brandon Munn, Joana Cabral, Rosalyn~J Moran,
  and Michael Breakspear.
\newblock Computational models link cellular mechanisms of neuromodulation to
  large-scale neural dynamics.
\newblock \emph{Nature neuroscience}, 24\penalty0 (6):\penalty0 765--776, 2021.

\bibitem[Singh et~al.(2023)Singh, van Breugel, Rao, and
  Brunton]{singh2023emergent}
Satpreet~H Singh, Floris van Breugel, Rajesh~PN Rao, and Bingni~W Brunton.
\newblock Emergent behaviour and neural dynamics in artificial agents tracking
  odour plumes.
\newblock \emph{Nature Machine Intelligence}, 5\penalty0 (1):\penalty0 58--70,
  2023.

\bibitem[Smola \& Sch{\"o}lkopf(1998)Smola and
  Sch{\"o}lkopf]{smola1998learning}
Alexander~J Smola and Bernhard Sch{\"o}lkopf.
\newblock \emph{Learning with kernels}, volume~4.
\newblock GMD-Forschungszentrum Informationstechnik Berlin, Germany, 1998.

\bibitem[Sohn et~al.(2019)Sohn, Narain, Meirhaeghe, and
  Jazayeri]{sohn2019bayesian}
Hansem Sohn, Devika Narain, Nicolas Meirhaeghe, and Mehrdad Jazayeri.
\newblock Bayesian computation through cortical latent dynamics.
\newblock \emph{Neuron}, 103\penalty0 (5):\penalty0 934--947, 2019.

\bibitem[Str{\"a}sser et~al.(2025)Str{\"a}sser, Worthmann, Mezić, Berberich,
  Schaller, and Allgöwer]{strasser2025overviewkoopmanbasedcontrolerror}
Robin Str{\"a}sser, Karl Worthmann, Igor Mezić, Julian Berberich, Manuel
  Schaller, and Frank Allgöwer.
\newblock An overview of koopman-based control: From error bounds to
  closed-loop guarantees, 2025.
\newblock URL \url{https://arxiv.org/abs/2509.02839}.

\bibitem[Sussillo et~al.(2015)Sussillo, Churchland, Kaufman, and
  Shenoy]{sussillo2015neural}
David Sussillo, Mark~M Churchland, Matthew~T Kaufman, and Krishna~V Shenoy.
\newblock A neural network that finds a naturalistic solution for the
  production of muscle activity.
\newblock \emph{Nature Neuroscience}, 18\penalty0 (7):\penalty0 1025--1033,
  2015.

\bibitem[Takeishi et~al.(2017{\natexlab{a}})Takeishi, Kawahara, and
  Yairi]{takeishi_learning_2017}
Naoya Takeishi, Yoshinobu Kawahara, and Takehisa Yairi.
\newblock Learning {Koopman} {Invariant} {Subspaces} for {Dynamic} {Mode}
  {Decomposition}.
\newblock In \emph{Advances in {Neural} {Information} {Processing} {Systems}},
  volume~30. Curran Associates, Inc., 2017{\natexlab{a}}.
\newblock URL
  \url{https://papers.nips.cc/paper/2017/hash/3a835d3215755c435ef4fe9965a3f2a0-Abstract.html}.

\bibitem[Takeishi et~al.(2017{\natexlab{b}})Takeishi, Kawahara, and
  Yairi]{takeishi_subspace_2017}
Naoya Takeishi, Yoshinobu Kawahara, and Takehisa Yairi.
\newblock Subspace dynamic mode decomposition for stochastic {Koopman}
  analysis.
\newblock \emph{Physical Review E}, 96\penalty0 (3):\penalty0 033310, September
  2017{\natexlab{b}}.
\newblock ISSN 2470-0045, 2470-0053.
\newblock \doi{10.1103/PhysRevE.96.033310}.
\newblock URL \url{https://link.aps.org/doi/10.1103/PhysRevE.96.033310}.

\bibitem[Van Der~Veen et~al.(2013)Van Der~Veen, Van~Wingerden, Bergamasco,
  Lovera, and Verhaegen]{van_der_veen_closedloop_2013}
Gijs Van Der~Veen, Jan‐Willem Van~Wingerden, Marco Bergamasco, Marco Lovera,
  and Michel Verhaegen.
\newblock Closed‐loop subspace identification methods: an overview.
\newblock \emph{IET Control Theory \& Applications}, 7\penalty0 (10):\penalty0
  1339--1358, July 2013.
\newblock ISSN 1751-8644, 1751-8652.
\newblock \doi{10.1049/iet-cta.2012.0653}.
\newblock URL
  \url{https://ietresearch.onlinelibrary.wiley.com/doi/10.1049/iet-cta.2012.0653}.

\bibitem[Van~Overschee \& De~Moor(1994)Van~Overschee and De~Moor]{van1994n4sid}
Peter Van~Overschee and Bart De~Moor.
\newblock N4sid: Subspace algorithms for the identification of combined
  deterministic-stochastic systems.
\newblock \emph{Automatica}, 30\penalty0 (1):\penalty0 75--93, 1994.

\bibitem[Verhaegen(1994)]{verhaegen1994identification}
Michel Verhaegen.
\newblock Identification of the deterministic part of mimo state space models
  given in innovations form from input-output data.
\newblock \emph{Automatica}, 30\penalty0 (1):\penalty0 61--74, 1994.

\bibitem[Verhaegen \& Verdult(2007)Verhaegen and
  Verdult]{Verhaegen_Verdult_2007}
Michel Verhaegen and Vincent Verdult.
\newblock \emph{Subspace model identification}, pp.\  292–344.
\newblock Cambridge University Press, 2007.

\bibitem[Vermani et~al.(2024)Vermani, Nassar, Jeon, Dowling, and
  Park]{VermaniNassarJeonDowlingPark2024MetaDynamicalStateSpace}
Ayesha Vermani, Josue Nassar, Hyungju Jeon, Matthew Dowling, and Il~Memming
  Park.
\newblock Meta-dynamical state space models for integrative neural data
  analysis.
\newblock \emph{arXiv preprint arXiv:2410.05454}, 2024.
\newblock revised 2025.

\bibitem[Versteeg et~al.(2025)Versteeg, McCart, Ostrow, Zoltowski, Washington,
  Driscoll, Codol, Michaels, Linderman, Sussillo,
  et~al.]{versteeg2025computation}
Christopher Versteeg, Jonathan~D McCart, Mitchell Ostrow, David~M Zoltowski,
  Clayton~B Washington, Laura Driscoll, Olivier Codol, Jonathan~A Michaels,
  Scott~W Linderman, David Sussillo, et~al.
\newblock Computation-through-dynamics benchmark: Simulated datasets and
  quality metrics for dynamical models of neural activity.
\newblock \emph{bioRxiv}, 2025.

\bibitem[Vinograd et~al.(2024)Vinograd, Nair, Kim, Linderman, and
  Anderson]{vinograd2024causal}
Amit Vinograd, Aditya Nair, Joseph~H. Kim, Scott~W. Linderman, and David~J.
  Anderson.
\newblock Causal evidence of a line attractor encoding an affective state.
\newblock \emph{Nature}, 634\penalty0 (8035):\penalty0 910--918, 2024.
\newblock \doi{10.1038/s41586-024-07915-x}.
\newblock URL \url{https://www.nature.com/articles/s41586-024-07915-x}.

\bibitem[Wang(1999)]{Wang1999SynapticPersistentActivity}
Xiao-Jing Wang.
\newblock Synaptic basis of cortical persistent activity: The importance of
  nmda receptors to working memory.
\newblock \emph{Journal of Neuroscience}, 19\penalty0 (21):\penalty0
  9587--9603, 1999.
\newblock \doi{10.1523/JNEUROSCI.19-21-09587.1999}.

\bibitem[Williams et~al.(2022)Williams, Kunz, Kornblith, and
  Linderman]{williams2022generalizedshapemetricsneural}
Alex~H. Williams, Erin Kunz, Simon Kornblith, and Scott~W. Linderman.
\newblock Generalized shape metrics on neural representations.
\newblock 2022.
\newblock URL \url{https://arxiv.org/abs/2110.14739}.

\bibitem[Williams et~al.(2016)Williams, Rowley, and
  Kevrekidis]{williams_kernel-based_2016}
Matthew~O. Williams, Clarence~W. Rowley, and Ioannis~G. Kevrekidis.
\newblock A kernel-based method for data-driven koopman spectral analysis.
\newblock \emph{Journal of Computational Dynamics}, 2\penalty0 (2):\penalty0
  247--265, May 2016.
\newblock ISSN 2158-2491.
\newblock \doi{10.3934/jcd.2015005}.
\newblock URL
  \url{https://www.aimsciences.org/article/doi/10.3934/jcd.2015005}.

\bibitem[Wu et~al.(2021)Wu, Brunton, and Revzen]{dmdchallenges}
Ziyou Wu, Steven~L. Brunton, and Shai Revzen.
\newblock Challenges in dynamic mode decomposition.
\newblock \emph{Journal of The Royal Society Interface}, 18\penalty0
  (185):\penalty0 20210686, 2021.
\newblock \doi{10.1098/rsif.2021.0686}.
\newblock URL
  \url{https://royalsocietypublishing.org/doi/abs/10.1098/rsif.2021.0686}.

\bibitem[Yamins et~al.(2014)Yamins, Hong, Cadieu, Solomon, Seibert, and
  DiCarlo]{yamins2014performance}
Daniel~LK Yamins, Ha~Hong, Charles~F Cadieu, Ethan~A Solomon, Darren Seibert,
  and James~J DiCarlo.
\newblock Performance-optimized hierarchical models predict neural responses in
  higher visual cortex.
\newblock \emph{Proceedings of the national academy of sciences}, 111\penalty0
  (23):\penalty0 8619--8624, 2014.

\end{thebibliography}
